\pdfoutput=1
\documentclass[aps,prd,print,preprintnumbers,showpacs]{revtex4}
\usepackage{latexsym,graphicx,amssymb,amsmath,mathrsfs}
\usepackage{setspace,bm}
\usepackage{hyperref}
\graphicspath{{./Figures/}}

\newcommand{\diff}{\mathrm{d}}
\newcommand{\p}{\partial}

\newcommand{\Diff}{{\mathcal{D}}}

\newcommand{\be}{\begin{equation}}      
\newcommand{\ee}{\end{equation}}      
\newcommand{\bea}{\begin{eqnarray}}      
\newcommand{\eea}{\end{eqnarray}}
\newcommand{\tn}{\widetilde{n}}
\newcommand{\tm}{\widetilde{m}}

\begin{document}

\preprint{RIKEN-QHP-114, RBRC-1062}

\title{ Phase structure of $SU(3)$ gauge-Higgs unification models at finite temperature}

\author{Kouji Kashiwa}
\email{kouji.kashiwa@yukawa.kyoto-u.ac.jp}
\affiliation{RIKEN/BNL, Brookhaven, National Laboratory, Upton, NY 11973}
\affiliation{Yukawa Institute for Theoretical Physics, Kyoto University, Kyoto 606-8502, Japan}

\author{Yuya Tanizaki}
\email{yuya.tanizaki@riken.jp}
\affiliation{Department of Physics, The University of Tokyo, Tokyo 113-0033, Japan}
\affiliation{Theoretical Research Division, Nishina Center, RIKEN, Wako 351-0198, Japan}

\date{\today}

\begin{abstract}
Five-dimensional $SU(3)$ gauge-Higgs unification models are studied at finite temperature in the warped extra dimension $S^1/\mathbb{Z}_2$. 
In order to investigate the phase structure, we develop a technique to compute the one-loop effective potential with the nontrivial Polyakov loop phase and with the nontrivial Wilson line phase along the extra dimension. 
Effective potentials as functions of two gauge-field condensations are shown for several simple matter contents, including fundamental, sextet, and adjoint representational Dirac fermions. 
Possible extensions and applications of our formalism are also briefly discussed. 
\end{abstract}


\maketitle

\section{Introduction}\label{sec:intro}
Spontaneous breaking of gauge symmetry is an important ingredient for our understanding of superconductivity in condensed matter physics and of electroweak interactions in the standard model of particle physics. 
In the standard model, the complex scalar field (Higgs field) acquires the vacuum expectation value and causes the electroweak symmetry breaking. 
However, there is a big theoretical question about the Higgs physics, the so-called fine-tuning problem: we still do not know why nature requires such an accurate cancellation between tree-level and quantum contributions to the mass of the Higgs particle so as to realize the 126 GeV light Higgs boson. 
If the standard model is realized as a low-energy effective theory of an ultraviolet complete theory, it is natural to assume some mechanism causing the accurate cancellation between those ultraviolet divergences. 

Gauge theories in higher-dimensional spacetime are candidates of such theories to go beyond the standard model. 
In gauge-Higgs unification, gauge bosons and Higgs fields are unified
into a five-dimensional gauge field, and then the Wilson line along the warped extra dimension $S^1/\mathbb{Z}_2$ behaves as the Higgs field in the four-dimensional spacetime~\cite{Manton:1979kb, Fairlie:1979zy, Fairlie:1979at, Hosotani:1983xw, Hosotani:1983jc, Hosotani:1988bm, Hatanaka:1998yp, Scrucca:2003ra}.
In this model, the mass of the Higgs particle as a five-dimensional local operator is prohibited due to the gauge symmetry in extra dimension, but the Higgs mass is realized as a nonlocal gauge-invariant quantity and generated dynamically through quantum correction. 

In this paper, we investigate the phase structure of $SU(3)$ gauge-Higgs unification models at finite temperature. 
In this case, not only the warped extra dimension but also the temporal dimension are compactified in the imaginary time formalism. Therefore, the Wilson lines along those two directions are candidates of gauge-invariant order parameters, and it is important to study effects of those two condensations of gauge fields. 
Indeed, in the case of strong interaction, the temporal Wilson loop, the so-called Polyakov loop, plays a central role in describing the confinement or deconfinement transition of quantum chromodynamics. On the other hand, the nontrivial Wilson line phase along the extra dimension causes the spontaneous gauge symmetry breaking, as mentioned above.

In order to understand the phase structure of gauge theories, the perturbative one-loop effective potential provides a good description in a weak-coupling region \cite{Gross:1981br,Weiss:1980rj}. 
Such calculation is nothing but the free-gas-limit calculation with the background gauge field, which is related to the order parameter of the gauge symmetry breaking. 
For our purpose, we must perform the computation of the one-loop effective potential with temporal and extra-dimensional background gauge fields. 
Therefore, we need to extend the previous formulation to evaluate the effective potential in the orbifold \cite{Kubo:2001zc, Haba:2004qf, Haba:2004qh, Panico:2005ft, Maru:2005jy, Sakamoto:2009hb} so as to also include the effect of the temporal gauge field. 

This paper is organized as follows. 
In Sec.~\ref{sec:formalism}, we first describe the basic formalism about the gauge-Higgs unification. In Sec.~\ref{sec:expression}, we derive the formula of the one-loop effective potential with two kinds of gauge field condensation, and show its analytic expression at finite temperature in the warped extra dimension. 
By introducing ultraviolet cutoff in three momenta, we can construct the effective potential for Wilson lines in the gauge-invariant way and discuss the effect of ultraviolet cutoff for physical quantities. 
Properties of the effective potential is shown in Sec.~\ref{sec:phase}. By adding simple matter contents to the five-dimensional pure gauge theory, we discuss the effect of Dirac fermions with fundamental, sextet, and adjoint representations in a systematic way. 
 Section~\ref{sec:summary} is devoted to summary. 
In Appendix~\ref{app:back}, we review the background field method briefly since it is useful in computation of the effective potential, and  the calculation for analytical expression of the effective potential is shown in detail in Appendix~\ref{app:cal}. Some useful formulas on the Lie algebra $\mathfrak{su}(3)$ are listed in Appendix~\ref{app:lie}.

\section{Basic formalism of Gauge-Higgs unification}\label{sec:formalism}

In this section, we describe the basic formalism of gauge-Higgs unification~\cite{Manton:1979kb, Fairlie:1979zy, Fairlie:1979at, Hosotani:1983xw, Hosotani:1983jc, Hosotani:1988bm, Hatanaka:1998yp, Scrucca:2003ra}. 
For that purpose, we consider a $SU(3)$ gauge theory on the five-dimensional spacetime $(S^1\times \mathbb{R}^3)\times S^1/\mathbb{Z}_2$ at finite temperature $T$, where $S^1$ is the temporal direction with perimeter $\beta=1/T$, $\mathbb{R}^3$ is the three-dimensional space, and $S^1/\mathbb{Z}_2$ is the underlying space of the orbifold. 
For constructing the orbifold, $\mathbb{Z}_2=\{\pm 1\}$ acts on the circle $S^1$ with radius $R$ as $y\cdot1 =y$ and $y\cdot (-1)=-y$ for $y\in S^1=\mathbb{R}/2\pi R\mathbb{Z}$. Under this action, there are two fixed points; $y=0,\pi R$, at which quantum fields obey a given boundary condition. 

\subsection{Gauge fields on orbifolds}
For the case of gauge fields, the boundary condition is given as
\bea
&A(\tau,\bm{x},y+2\pi R)= U A(\tau,\bm{x},y) U^{\dagger},\\
&\left\{\begin{array}{c}
A_{\mu}(\tau,\bm{x},-y)= P_0 A_{\mu}(\tau,\bm{x},y) P_0^{\dagger}\\
A_{y}(\tau,\bm{x},-y)=- P_0 A_{y}(\tau,\bm{x},y) P_{0}^{\dagger}
\end{array}\right.\\
&\left\{\begin{array}{c}
A_{\mu}(\tau,\bm{x},\pi R-y)= P_1 A_{\mu}(\tau,\bm{x},\pi R+y) P_1^{\dagger}\\
A_{y}(\tau,\bm{x},\pi R-y)=- P_1 A_{y}(\tau,\bm{x},\pi R+y) P_{1}^{\dagger}
\end{array}\right.
\eea
with $U=U^{\dagger}$, $P_i=P_i^{\dagger}=P_i^{-1}$, and the consistency condition implies that $U=P_1P_0$. 
Due to this boundary condition, the $SU(3)$ gauge symmetry in the five-dimensional spacetime can be explicitly broken at these boundaries. However, there still exists the $SU(3)$ gauge symmetry inside the bulk, and thus the mass of gauge fields is prohibited. 
In order to describe the electroweak theory, $SU(3)$ is explicitly broken to $SU(2)\times U(1)$ by choosing $P\equiv P_i=\mathrm{diag}(-1,-1,1)$. 

Gell-Mann matrices $T^a$ satisfy $P T^a P^{\dagger}=T^a$ for $a=1,2,3,$ and $8$, and $P T^a P^{\dagger} =-T^a$ for $a=4,\ldots,7$. Using the matrix notation given in Appendix \ref{app:lie}, quantum numbers of each field under the parity $P$ in the extra dimension are given as 
\bea
P(A_{\mu})=\left(\begin{array}{cc|c}+&+&-\\+&+&-\\\hline-&-&+\end{array}\right),\qquad 
P(A_{y})=\left(\begin{array}{cc|c}-&-&+\\-&-&+\\\hline+&+&-\end{array}\right), 
\eea
for $\mu=0,\ldots,3$. 
Since fields with negative parity cannot take any nonzero constant values, the Kaluza-Klein zero modes of the $SU(3)$ gauge field $A(x,y)$ are decomposed into the $SU(2)$-gauge field $A_{\mu}^{1,2,3}(x)$, the $U(1)$-gauge field $A^8_{\mu}(x)$, and the matter fields $A_{y}^{4,5,6,7}$ with $SU(2)$-(anti)fundamental representation. 
The electroweak theory as a low-energy effective theory consists of the $SU(2)$ gauge field $A_{\mu}^{1,2,3}$, the $U(1)$ gauge field $A^{8}_{\mu}$, and the Higgs field $\Phi\sim \left(\begin{array}{c}A^{4}_y-iA^{5}_y\\ A^{6}_y-iA^{7}_y\end{array}\right)$ with its complex conjugate. 

We consider two different kinds of condensate $\langle A_0 \rangle={\pi
T\over g}a_0$ and
$\langle A_y\rangle={1\over gR}a_y$ with the five-dimensional gauge coupling $g$.
Then the classical potential for these condensates is
\be
{1\over g^2}\mathrm{Tr}\left([a_0,a_y]^2\right). 
\ee
As long as the five-dimensional gauge coupling is sufficiently small, this term must be minimized at the leading order. Within this approximation, it is good to assume that 
\be
[a_0 , a_y] =0.   
\ee
As a result, the field strength of this background field vanishes. In the following discussion, we always use this approximation. 

The spontaneous symmetry breaking of the electroweak theory is parametrized by the vacuum expectation value of the Higgs field $\Phi$. In gauge-Higgs unification models, it is equivalent to set 
\be
\langle A_y\rangle = {a\over g R}T^6, 
\label{eq:background}
\ee
with $a$ a real number. 
From the previous constraint, the $A_0$ condensate must be proportional to $T^3+T^8/\sqrt{3}$ so that  
\be
\langle A_0\rangle={2\pi T\over g}q{3(T^3+T^8/\sqrt{3})\over 2}
={2\pi T\over g}{q}\left(\begin{array}{ccc}
1&0&0\\
0&-1/2&0\\
0&0&-1/2
\end{array}\right).
\label{eq:background_temporal}
\ee
This basis makes clear the explicit breaking of $SU(3)$ into $SU(2)\times U(1)$ due to the $A_y^6$ condensation. 
Therefore, we chose the basis of the Cartan subalgebra by $H_1=T^6$ and $H_2={\sqrt{3}\over 2}(T^3+T^8/\sqrt{3})$. 
Details about the Lie algebra $\mathfrak{su}(3)$ are given in Appendix~\ref{app:lie}. 

These expectation values must be regarded as phases of corresponding Wilson loops. With the background field (\ref{eq:background}), the Wilson line along the extra dimension $W$ is given by 
\be
W
 = {\cal P} \exp \Bigl( ig \int_0^{2\pi R} dy \langle A_y \rangle  \Bigr)
= \left(
    \begin{array}{ccc}
      1 & 0                   & 0 \\
      0 &  \cos( \pi a) & i\sin( \pi a) \\
      0 & i\sin( \pi a) &  \cos( \pi a)
    \end{array}
  \right). 
\label{eq:wilson_extra}
\ee
Here, ${\cal P}$ refers to the path-ordering operator. We should notice that there is an identification $a\sim a+2$, since $a$ is the Wilson line phase along the extra dimension, and any local potential of $a$ is forbidden due to this remnant of the gauge symmetry. Therefore, potentials of $a$ must be nonlocal in the extra dimension.  Depending on the vacuum expectation values of the Wilson line $W$, patterns of the gauge symmetry breaking are very different: 
\be
SU(3)\xrightarrow{\mathrm{orbifolding}}SU(2)\times U(1)\xrightarrow{\mathrm{SSB}}\left\{
\begin{array}{cc}
SU(2)\times U(1) & a=0\\
U(1)\times U(1) & a=1\\
U(1)& \mathrm{otherwise}
\end{array}
\right. . 
\ee
On the other hand, condensation of $A_0$ is related to the Polyakov loop, the Wilson line along the temporal direction, at the classical level:  
\be
P={\cal P}\exp\left(ig \int_0^{\beta}\diff \tau \langle A_0\rangle\right)
=\mathrm{diag}\left(\exp i{2\pi q },\exp{-i{\pi q}},\exp{-i{\pi q}}\right). 
\label{eq:polyakov}
\ee
Again, we have an identification $q\sim q+2$. Due to this gauge symmetry, potentials of $q$ must also be nonlocal along the temporal direction.  When $q=0$, ${2\over 3}$, and ${4\over 3}$, the Polyakov loops are proportional to the unit matrix, which are center elements of the gauge group $SU(3)$.  

\subsection{Fermions on orbifolds}
We consider Dirac fermions $\Psi$ on the five-dimensional spacetime with a certain representation $\mathcal{R}$ of the gauge group $SU(3)$. The kinetic term is given by 
\be
\mathcal{L}=\overline{\Psi}\gamma^I \left(\p_I +ig \mathcal{R}(A_I)\right)\Psi. 
\ee
Under the parity operation $y\mapsto -y$, the Dirac fermion must obey the transformation 
\be
\psi(x,-y)=\eta \mathcal{R}(P) \gamma^5 \psi(x,y),\; \psi(x,\pi R-y)=\eta' \mathcal{R}(P) \gamma^5 \psi(x,\pi R+y). 
\ee
with $P=\mathrm{diag}(-1,-1,1)\in SU(3)$. Here, $\eta$ and $\eta'$ are parameters, which take $+1$ or $-1$. 
For the Dirac fermion, left and right particles obey the opposite intrinsic parity also in the warped extra dimension: 
\bea
&&\psi_L(x,-y)=\eta \mathcal{R}(P)\psi_L(x,y),\; \psi_R(x,-y)=-\eta \mathcal{R}(P)\psi_R(x,y),\nonumber\\ 
&&\psi_L(x,\pi R-y)=\eta' \mathcal{R}(P)\psi_L(x,\pi R+y),\; \psi_R(x,\pi R-y)=-\eta' \mathcal{R}(P)\psi_R(x,\pi R+y).
\label{eq:dirac_left_right} 
\eea
The consistency condition again implies that 
\be
\psi(x,y+2\pi R)=\eta\eta' \psi(x,y).
\ee 
Depending on the value of the product $\eta\eta'=\pm1$, the Dirac fermion obeys the periodic or antiperiodic boundary condition in terms of $y$, respectively.

\section{One-loop effective potential with two condensations}\label{sec:expression}
In this section, we compute one-loop effective potential for five-dimensional $SU(3)$ gauge theories with two different kinds of condensation (\ref{eq:background}) and (\ref{eq:background_temporal}). 
For that purpose, we extend the previous formalism to calculate the one-loop effective potential on the orbifold
\cite{Kubo:2001zc, Haba:2004qf, Haba:2004qh, Panico:2005ft, Maru:2005jy, Maru:2006wx,  Sakamoto:2009hb}.

\subsection{Formula of the effective potential for gauge fields}
The one-loop effective potential of gauge fields $\mathcal{V}_{\mathrm{eff}}^{\mathrm{g+gh}}$ is defined by 
\be
\mathcal{V}_{\mathrm{eff}}^{\mathrm{g+gh}}=-\ln \int \Diff A_I \Diff \overline{c}\Diff c \exp\left(-\int\diff^4 x\diff y \; \mathrm{Tr}\left[ (D^{\mathrm{cl}}_IA_J)^2+\overline{c}D^{\mathrm{cl}}_ID^{\mathrm{cl}}_I c \right]\right), 
\ee
where $D^{\mathrm{cl}}_I A_J=\p_I A_J+ig [\langle A_I\rangle, A_J]$ with $I,J=0,\ldots,4$. Here, the background field gauge is chosen, and detailed derivation is given in Appendix \ref{app:back}. 

In order to calculate this functional integration, we consider the eigenvalue problem of the operator $-(D^{\mathrm{cl}}_I)^2$. Let $A(x,y)$ be an adjoint representational field, which satisfies the boundary condition $A(x,-y)=PA(x,y)P^{\dagger}$ and $A(x,y+2\pi R)=A(x,y)$. 
In the following, we introduce the mass scale $M=1/2\pi R$ determined by the size of the extra dimension. 
For $a=1,2,3,$ and $8$, $A^a(x,y)$ can be decomposed into the Kaluza-Klein modes \cite{Kaluza:1921tu,Klein:1926tv} as 
\be
A^a(x,y)=\sqrt{M}A^a_0(x)+\sqrt{2M}\sum_{m=1}^{\infty}A^a_m(x)\cos\left(2\pi Mm y\right), \label{eq:positive_parity}
\ee
and, for $b=4,5,6$, and $7$, 
\be
A^b(x,y)=\sqrt{2M}\sum_{m=1}^{\infty}A^b_m(x)\sin\left(2\pi Mm y\right). \label{eq:negative_parity}
\ee
The important difference between (\ref{eq:positive_parity}) and (\ref{eq:negative_parity}) is the existence of Kaluza-Klein zero modes. Each mode $A^a_m(x)$ can be written as the summation over Matsubara modes: 
\be
A^a_m(x)=\sqrt{T}\sum_{n=-\infty}^{\infty}A^a_{n,m}(\bm{x})e^{2\pi i Tn\tau}. 
\ee
Using this decomposition, it suffices to evaluate the quadratic form $\int \diff^4 x\diff y\mathrm{Tr}(D^{\mathrm{cl}}_{I}A)^2$ for solving the eigenvalue problem. The integration over the warped extra dimension gives 
\bea
&&\int\diff y \sum_{I=0}^{4}\mathrm{Tr}(D^{\mathrm{cl}}_I A(x,y))^2=\sum_{I=0}^{4}\mathrm{Tr}\left(D^{\mathrm{cl}}_I \sum_{a=1,2,3,8}A_0^a(x)T^a\right)^2\nonumber\\
&&+\sum_{m=1}^{\infty}\left[\sum_{\mu=0}^{3}\mathrm{Tr}(D^{\mathrm{cl}}_\mu A_m(x))^2+\mathrm{Tr}\left(-(2\pi M m )PA_m(x)P^{\dagger}+ig[\langle A_y\rangle,A_m(x)]\right)^2\right], 
\eea
where $A_m(x)=\sum_{a=1}^{8}A^a_m(x) T^a$. The parity operation in the last term comes from the difference of the basis, and the cross term of the squared does not vanish. This comes from the fact that the $y$ derivative $\p_y$ and $[\langle A_y\rangle,\cdot](\propto [H_1,\cdot])$ transform in the same way under the parity $y\mapsto -y$. 
By substituting (\ref{eq:background}), the last term can be written as $(m\ge 1)$
\bea
&&\mathrm{Tr}\left(-(2\pi M m )PA_m(x)P^{\dagger}+ig[\langle A_y\rangle,A_m(x)]\right)^2\nonumber\\
&=&{(2\pi M)^2\over 2}\left[(A^1_m\; A^5_m)\left(\begin{array}{cc}m^2+\left({a\over 2}\right)^2&ma\\ma&m^2+\left({a\over 2}\right)^2\end{array}\right)\left(\begin{array}{c}A^1_m\\A^5_m\end{array}\right)\right.
+(A^2_m\; A^4_m)\left(\begin{array}{cc}m^2+\left( {a\over 2}\right)^2 & -ma \\ -ma & m^2+\left({a\over 2}\right)^2 \end{array}\right)\left(\begin{array}{c}A^2_m\\A^4_m\end{array}\right)\nonumber\\
&&+m^2 (A^6_m(x))^2+\left. (A^3_m\; A^7_m\; A^8_m) \left(\begin{array}{ccc} m^2+\left({a\over 2}\right)^2 & -ma & -{\sqrt{3}\over 4}a^2 \\-m a & m^2+a^2 & \sqrt{3}ma \\ -{\sqrt{3}\over 4}a^2 & \sqrt{3}ma & m^2+\left({\sqrt{3}a\over2}\right)^2 \end{array}\right) \left(\begin{array}{c} A^3_m \\ A^7_m \\ A^8_m \end{array}\right) \right]. \label{eq:y_comp}
\eea
Some useful formulas on the Lie algebra $\mathfrak{su}(3)$ in this computation are shown in Appendix~\ref{app:lie}. 
The important point of this expression is that the condensation of $A_6$ mixes the odd and even parity states. 
Since $H_2={\sqrt{3}\over 2}(T^3+T^8/\sqrt{3})$ only mixes $A^1_m$ and $A^2_m$ and also $A^4_m$ and $A^5_m$, diagonalization of $H_2$ does not affect that of the quadratic form (\ref{eq:y_comp}).  
As a result, the one-loop effective potential of an adjoint representational field is given as 
\bea
\mathcal{V}_{\mathrm{eff}}^{\mathrm{adj}}&=&{1\over 2}\int{\diff^3\bm{p}\over (2\pi)^3}TM\sum_{n,m=-\infty}^{\infty}\Bigl[\ln\left((2\pi T)^2 n^2+\bm{p}^2+(2\pi M)^2 m^2 \right)\nonumber\\
&&+2\ln\left((2\pi T)^2\left(n+{3q\over 2}\right)^2+\bm{p}^2+(2\pi M)^2\left(m+{a\over 2}\right)^2\right)\nonumber\\
&&+\ln\left((2\pi T)^2n^2+\bm{p}^2+(2\pi M)^2\left(m+a\right)^2\right)\Bigr]. 
\label{eq:energy_adjoint_rep}
\eea
Now we can readily see that the same expression for the effective potential is obtained even from an adjoint representational field with the opposite parity. Therefore, the gauge field contribution to the effective potential is given by 
\be
\mathcal{V}_{\mathrm{eff}}^{\mathrm{g+gh}}=3\mathcal{V}_{\mathrm{eff}}^{\mathrm{adj}}. 
\ee

\subsection{Formula of the effective potential for matter fields}
\begin{figure}[t]
\centering
\includegraphics[width=4.4cm]{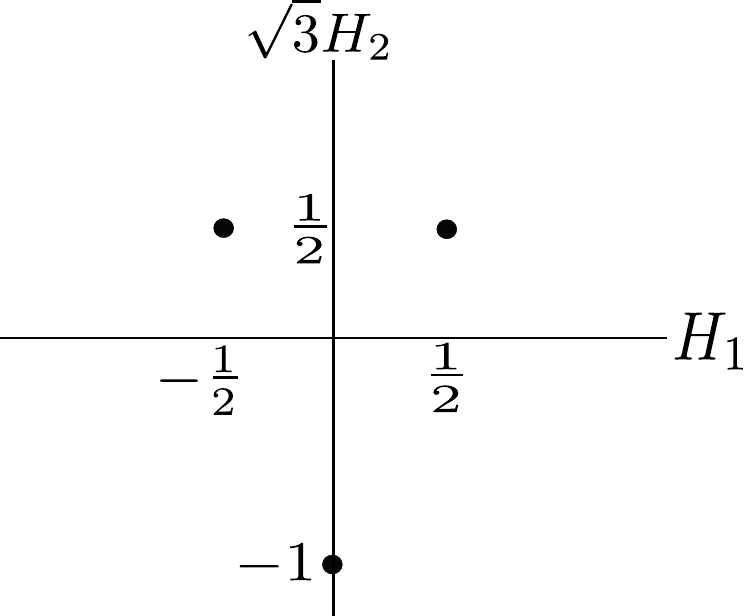}\hspace{1.4cm} 
\includegraphics[width=3.2cm]{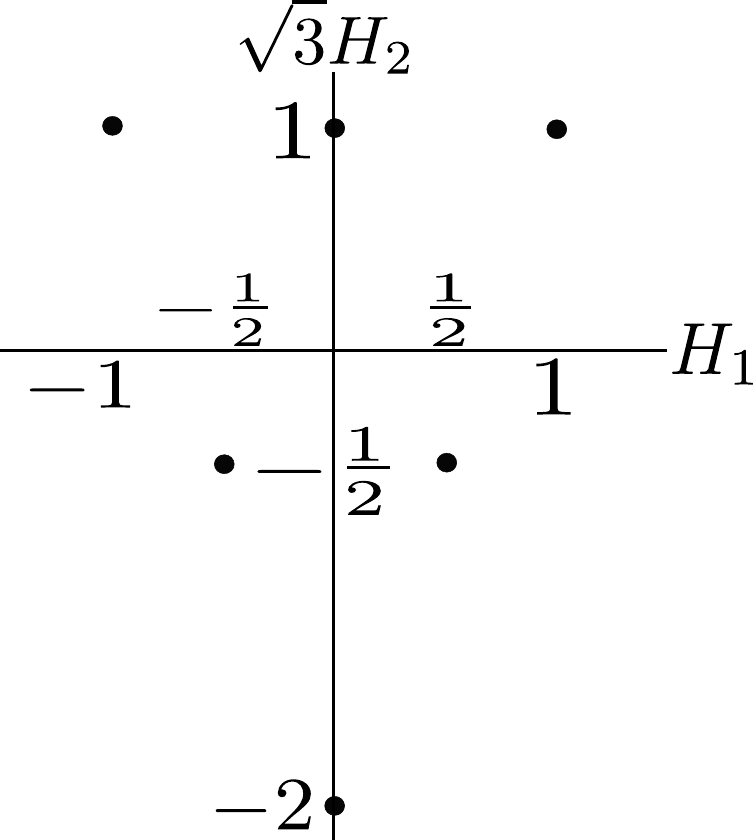}\hspace{1.4cm}
\includegraphics[width=3.2cm]{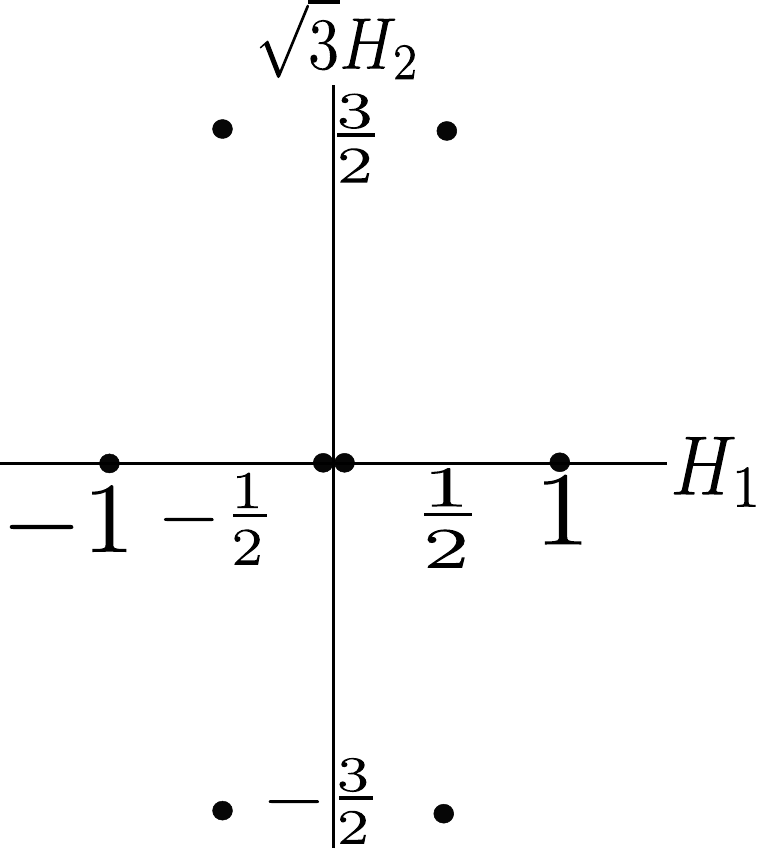}
\caption{Weight diagrams of the three, six, and eight-dimensional representations with $H_1=T^6$ and $H_2={\sqrt{3}\over 2}(T^3+T^8/\sqrt{3})$. }
\label{fig:weight}
\end{figure}

Let us consider the contribution of matters to the one-loop effective potential $\mathcal{V}_{\mathrm{eff}}$.  Fundamental, sextet, and adjoint Dirac fermions will be considered as matter contents, and weight
diagrams of those representations are given in Fig. \ref{fig:weight}. 
For simplicity of computation, let us specify $\eta\eta'=+1$ for matter fields in the following computation. Extension to the case $\eta\eta'=-1$ is straightforward. 

For the fundamental representational matter field, two states with weight vectors $(\pm 1/2,1/2\sqrt{3})$ are linear combinations of opposite-parity states, and the last one with $(0,-1/\sqrt{3})$ has a definite parity $\eta(=\eta')$. Then, the summation over Kaluza-Klein modes for the one-loop effective potential becomes 
\bea
&&{1+\eta\over 2}M\ln\left[(2\pi T)^2 \left(n+{q}\right)^2+\bm{p}^2\right] + M \sum_{m=1}^{\infty} \ln\left[(2\pi T)^2\left(n+{q}\right)^2+\bm{p}^2 + (2\pi M)^2 m^2 \right]\nonumber\\
&&+M\sum_{m=-\infty}^{\infty}\ln\left[(2\pi T)^2\left(n+{q\over 2}\right)^2+\bm{p}^2+(2\pi M)^2\left(m+{a\over 2}\right)^2\right]. 
\eea
Therefore, we obtain the following expression for the effective potential of a single fundamental representational field: 
\bea
&&{1\over 2}\int{\diff^3\bm{p}\over(2\pi)^3}T\sum_{n=-\infty}^{\infty}{\eta\over 2} \ln\left[(2\pi T)^2\left(n+q+{F\over 2}\right)^2+\bm{p}^2\right]\nonumber\\
&&+{1\over 2}\int{\diff^3\bm{p}\over(2\pi)^3}TM\sum_{n,m=-\infty}^{\infty} \left[ {1\over 2}\ln\left\{(2\pi T)^2\left(n+{q}+{F\over 2}\right)^2+\bm{p}^2+(2\pi M)^2 m^2\right\}\right.\nonumber\\
&&\left.+\ln\left\{(2\pi T)^2\left(n+{q\over 2}+{F\over 2}\right)^2+\bm{p}^2+(2\pi M)^2 \left(m+{a\over 2} \right)^2\right\}\right],
\label{eq:expression_fundamental}
\eea
where $F=+1$ for fermions and $F=0$ for bosons.  
In the case of Dirac fermions, left and right particles have opposite intrinsic parity $\eta$ and $-\eta$ according to (\ref{eq:dirac_left_right}), and thus the first term cancels by summing up all of them. Therefore, each fundamental Dirac fermion with $\eta\eta'=+1$  contributes to the effective potential as 
\bea
\mathcal{V}_{\mathrm{eff}}^{\mathrm{fd}}&=&4\times {1\over 2}\int{\diff^3\bm{p}\over(2\pi)^3}TM\sum_{n,m=-\infty}^{\infty} \left[ {1\over 2}\ln\left\{(2\pi T)^2\left(n+{q}+{1\over 2}\right)^2+\bm{p}^2+(2\pi M)^2 m^2\right\}\right.\nonumber\\
&&\left.+\ln\left\{(2\pi T)^2\left(n+{q\over 2}+{1\over 2}\right)^2+\bm{p}^2+(2\pi M)^2 \left(m+{a\over 2} \right)^2\right\}\right]. 
\eea

Similarly, the contribution of the sextet-representational Dirac fermion with $\eta\eta'=+1$ is given by 
\bea
\mathcal{V}_{\mathrm{eff}}^{\mathrm{sxt}}&=&4\times {1\over 2}\int{\diff^3\bm{p}\over(2\pi)^3}TM\sum_{n,m=-\infty}^{\infty} 
\left[ {1\over 2}\ln\left\{(2\pi T)^2\left(n+{q}+{1\over 2}\right)^2+\bm{p}^2+(2\pi M)^2 m^2\right\}\right.\nonumber\\
&&+{1\over 2}\ln\left\{(2\pi T)^2\left(n+{2q}+{1\over 2}\right)^2+\bm{p}^2+(2\pi M)^2 m^2\right\}\nonumber\\
&&+\ln\left\{(2\pi T)^2\left(n+{q\over 2}+{1\over 2}\right)^2+\bm{p}^2+(2\pi M)^2 \left(m+{a\over 2}\right)^2\right\}\nonumber\\
&&\left.+\ln\left\{(2\pi T)^2\left(n+{q}+{1\over 2}\right)^2+\bm{p}^2+(2\pi M)^2 \left(m+{a} \right)^2\right\}\right]. 
\eea

Expression for the adjoint representational field can be obtained in the same way presented in the previous subsection. 

\subsection{Analytic formula of the effective potential}
So far in this section, we have derived the naive expression for the one-loop effective potential. 
However, the five-dimensional gauge theory is nonrenormalizable and thus it is natural to introduce some ultraviolet (UV) cutoff to the theory. 
In order to respect discrete shift symmetries $a\mapsto a+2$ and $q\mapsto q+2$ of the effective potential, we introduce the UV cutoff in the three spatial momentum integration. 
The UV cutoff effect to the Higgs mass at zero temperature is investigated for the $SU(3)$ gauge-Higgs unification model with a four-dimensional momentum cutoff in Ref.~\cite{So:2013cy}.  

In order to derive the analytic expression for the perturbative one-loop effective potential, we need to calculate 
\be
F_T(q,a)={1\over 2}\int_{0}^{\Lambda}{\diff^3\bm{p}\over (2\pi)^3}TM\sum_{n,m=-\infty}^{\infty}
\ln\left[{\bm{p}^2+(2\pi T)^2(n+{q\over 2})^2+(2\pi M)^2 (m+{a\over 2})^2 \over \bm{p}^2+(2\pi T)^2 n^2+(2\pi M)^2 m^2}\right],  \label{eq:free_energy}
\ee
with $M=1/2\pi R$. Here, $\Lambda$ is the UV cutoff in the three spatial momentum. Since we are only interested in its field dependence, field independent parts of the free energy are disregarded in the following calculation. The UV cutoff dependence of the free energy can be extracted according to the following decomposition: 
\be
F_{T}( q ,a)=F_T( q ,a)|_{\Lambda\mathrm{-indep.}}+F_T( q ,a)|_{\Lambda\mathrm{-dep.}}. 
\ee
The first term is defined by the limit $\Lambda\to \infty$ of the field-dependent part of (\ref{eq:free_energy}), and its expression is given by 
\be
F_T( q ,a)|_{\Lambda\mathrm{-indep.}}=-{3M^5\over 4\pi^2}\left[\sum_{\tm\ge 1}{\cos \pi \tm a\over \tm^5}+\sum_{\tn\ge 1}{\cos \pi \tn  q \over \sqrt{{M^2\over T^2}\tn^2}^5}
+\sum_{\tm,\tn\ge 1}{2\cos \pi \tn  q  \cos \pi \tm a\over \left(\tm^2+{M^2\over T^2}\tn^2\right)^{5/2}}\right]. 
\ee
The cutoff dependence comes from the second term, 
\bea
F_T(q,a)|_{\Lambda\mathrm{-dep.}}&=&-{M^5\over \pi^3}\sum_{\tm\ge 1}{\cos \pi \tm a\over \tm^5}\widetilde{G}\left({\Lambda\over 2M}\tm\right)
-{M^5\over \pi^3}\sum_{\tn\ge 1}{\cos \pi \tn  q \over \sqrt{{M^2\over T^2}\tn^2}^5}\widetilde{G}\left({\Lambda\over 2T}\tn\right) \nonumber\\
&&-{2M^5\over \pi^3}\sum_{\tm,\tn\ge 1}{\cos \pi \tn  q  \cos \pi \tm a\over \left(\tm^2+{M^2\over T^2}\tn^2\right)^{5/2}}\widetilde{G}\left({\Lambda\over 2M}\sqrt{\tm^2+{M^2\over T^2}\tn^2}\right), 
\eea
where $\widetilde{G}$ is defined using the Meijer $G$ function and the
modified Bessel function of the second kind as 
\be
\widetilde{G}(z)=G^{2,1}_{1,3}\left(\begin{array}{ccc}&1&\\{1\over 2}&{5\over 2}&0\end{array};z^2\right)-{3\pi \over 4}-4z^3 K_2(2z). 
\ee
Detailed derivation is given in Appendix~\ref{app:cal}, and we neglected the field-independent part which is not of our interest. We also show in Appendix~\ref{app:cal} that UV cutoff dependence disappears if $\Lambda\gtrsim 10 \mathrm{max}\{M,T\}$, which is consistent with the result in Ref.~\cite{So:2013cy}. 
Therefore, for a good predictability of this cutoff theory on the properties of Wilson line operators, the Kaluza-Klein mass scale and temperatures must be smaller than $\Lambda/10$. 
In the following, we assume that the UV cutoff of three-momentum satisfies this condition, and we extract the finite effective potential for background gauge fields $a$ and $q$. 

According to Fig.~\ref{fig:weight}, the effective potentials for fundamental, sextet, and adjoint representations are given as 
\bea
F^{\mathrm{fd}}_T(q,a,F,\delta)&=&F_T(q+F,a+\delta)+{1\over 2}F_T(2q+F,\delta) ,\\
F^{\mathrm{\bm{sxt}}}_T(q,a,F,\delta)&=&{1\over 2}F_T(2q+F,\delta) +{1\over 2} F_T(4q+F,\delta)+F_T(q+F,a+\delta)+F_T(2q+F,2a+\delta), \\
F^{\mathrm{adj}}_T(q,a,F,\delta)&=&2 F_T(3q+F,a+\delta)+F_T(F,2a+\delta),  
\eea
respectively. Here, $F$ represents the fermion number of the field to be considered, and $\delta$ takes the values $0$ and $1$ for $\eta\eta'=+1$ and $-1$, respectively. 
Here, we implicitly assume that matter fields form a multiplet so as to erase terms like the first term in (\ref{eq:expression_fundamental}), and we must emphasize that Dirac fermions satisfy this condition. 
Using these quantities, the general expression of the effective potential can be written in the following way: 
\bea
{\cal V}_{\mathrm{eff}}&=&3F_T^{\mathrm{adj}}(q,a,0,0)\nonumber\\
&-&4N_F^{\mathrm{adj}(+)}F^{\mathrm{adj}}_T(q,a,1,0)-4N_F^{\mathrm{sxt}(+)}F^{\mathrm{sxt}}_T(q,a,1,0)-4N_F^{\mathrm{fd}(+)}F^{\mathrm{fd}}_T(q,a,1,0) \nonumber\\
&-&4N_F^{\mathrm{adj}(-)}F^{\mathrm{adj}}_T(q,a,1,1)-4N_F^{\mathrm{sxt}(-)}F^{\mathrm{sxt}}_T(q,a,1,1)-4N_F^{\mathrm{fd}(-)}F^{\mathrm{fd}}_T(q,a,1,1).
\eea
Here $N_F^{\mathrm{fd}(\pm)}$, $N_F^{\mathrm{sxt}(\pm)}$, and $N_F^{\mathrm{adj}(\pm)}$ represent the number of Dirac fermions with fundamental, sextet, and adjoint representation with $\eta\eta'=\pm1$, respectively. 


\section{Phase structure of five-dimensional $SU(3)$ gauge theories}\label{sec:phase}
In this section, we discuss phase structure of five-dimensional $SU(3)$ gauge theories in a systematic way motivated by electroweak phase transition in the gauge-Higgs unification. 
Since there is an identification $a\sim a+2$ and $q\sim q+2$, and a symmetry $a\mapsto -a$ and $q\mapsto -q$, it suffices to show behaviors of the effective potential in the region $0\le a,q\le 1$. 

\begin{figure}[t]
\centering
\includegraphics[scale=.5]{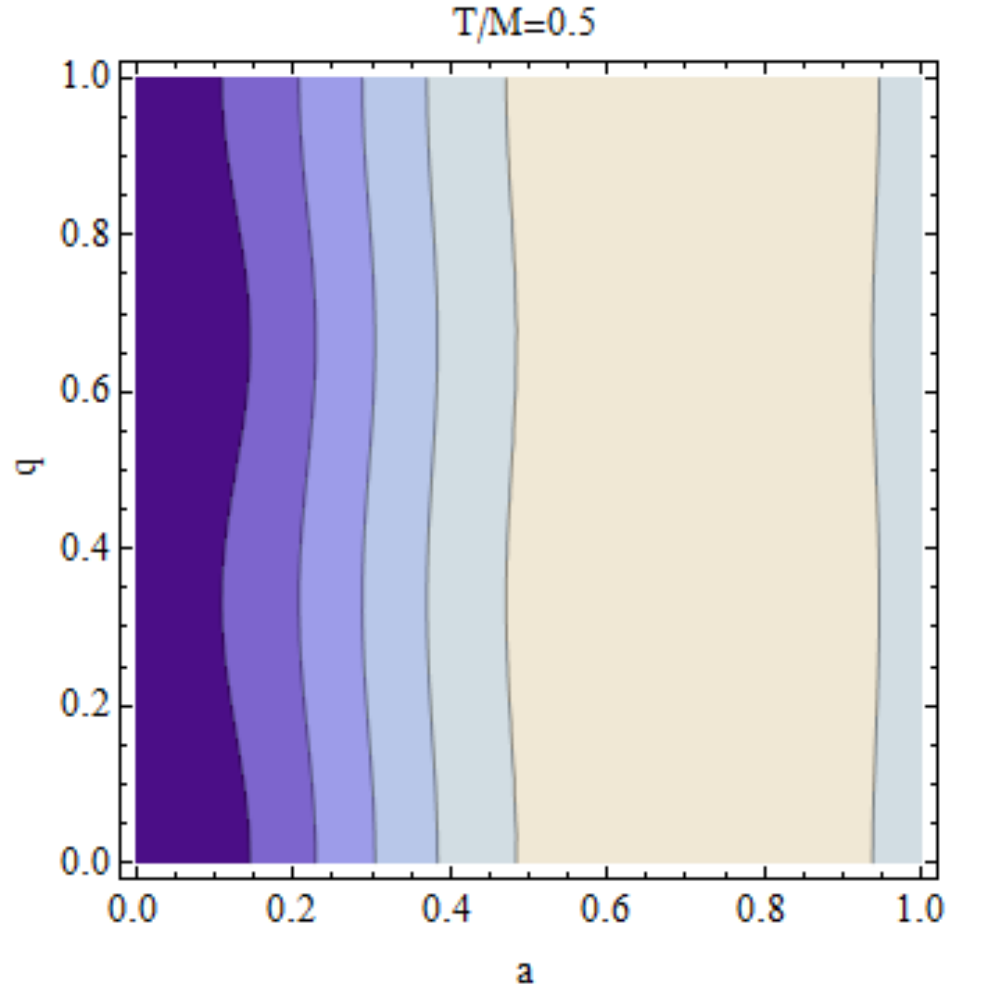}\hspace{.5cm}
\includegraphics[scale=.5]{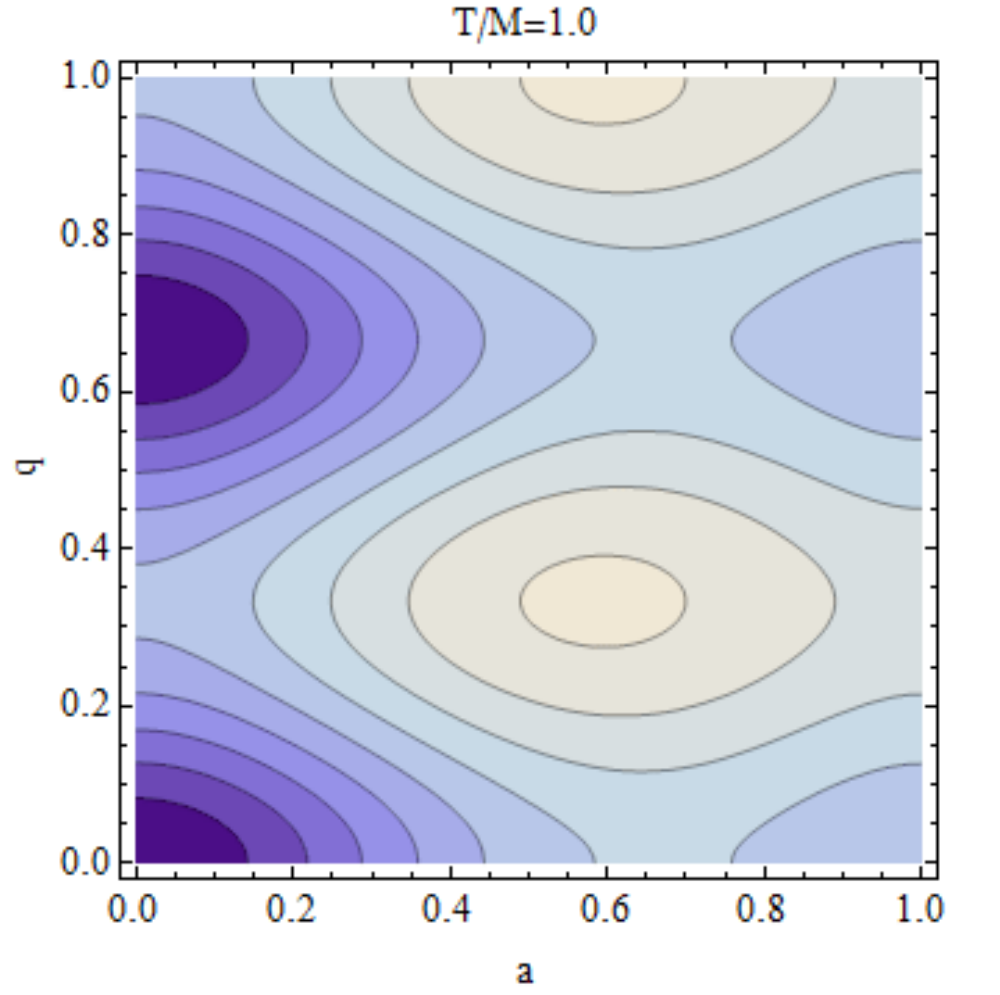}\hspace{.5cm}
\includegraphics[scale=.5]{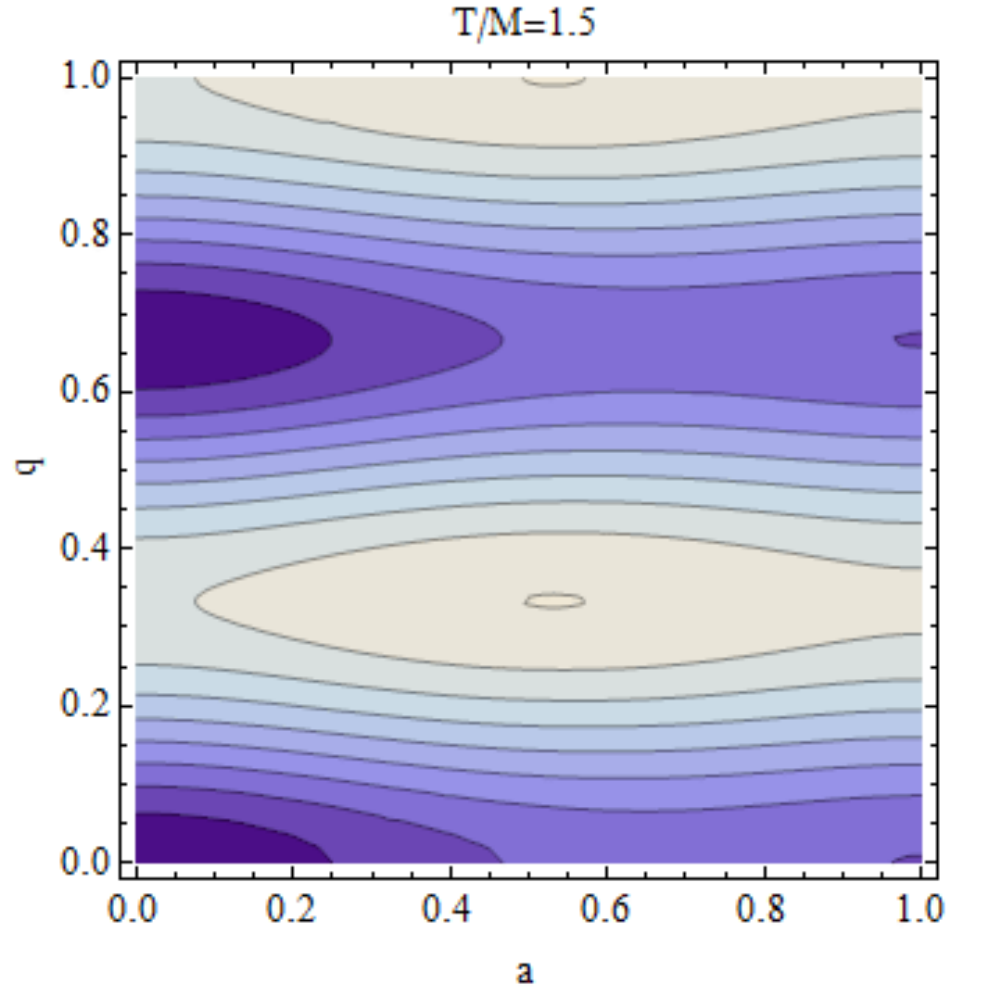}
\caption{Contour plots of the effective potential (\ref{eq:eff_pot_gauge}) only with gauge bosons in terms of $a$ and $q$ at $T/M=0.5,1.0,1.5$. As a result of the center symmetry, the figure is symmetric under $q\mapsto q+{2\over 3}$.}
\label{fig:eff_pot_gauge}
\end{figure}

As a first step, let us start with the theory only with gauge fields. In this case, the effective potential is given by 
\be
{\cal V}_{\mathrm{eff}}(a,q)=3F_T^{\mathrm{adj}}(q,a,0,0). 
\label{eq:eff_pot_gauge}
\ee
Contour plots of this effective potential at $T/M=0.5$, $1.0$, and $1.5$
are shown in Fig.~\ref{fig:eff_pot_gauge}. According to these plots, we can observe that there always exist degenerate vacua at $(a,q)=(0,0)$, $(0,2/3)$, and $(0,4/3)$. These three minima are connected by the center symmetry $\mathbb{Z}_3$ of the five-dimensional gauge group $SU(3)$. 
This degeneracy can be explicitly seen from the expression (\ref{eq:energy_adjoint_rep}): the effective potential has a center symmetry $q\mapsto q+2/3$. 

There is a stripe along $q$ direction when temperature $T$ is much
smaller than the Kaluza-Klein mass scale $M$ (see the case $T/M=0.5$ of
Fig.~\ref{fig:eff_pot_gauge}). This is a common feature for the effective potentials, since they become independent of $q$ at sufficiently low temperatures.

\begin{figure}[t]
\centering
\includegraphics[scale=.5]{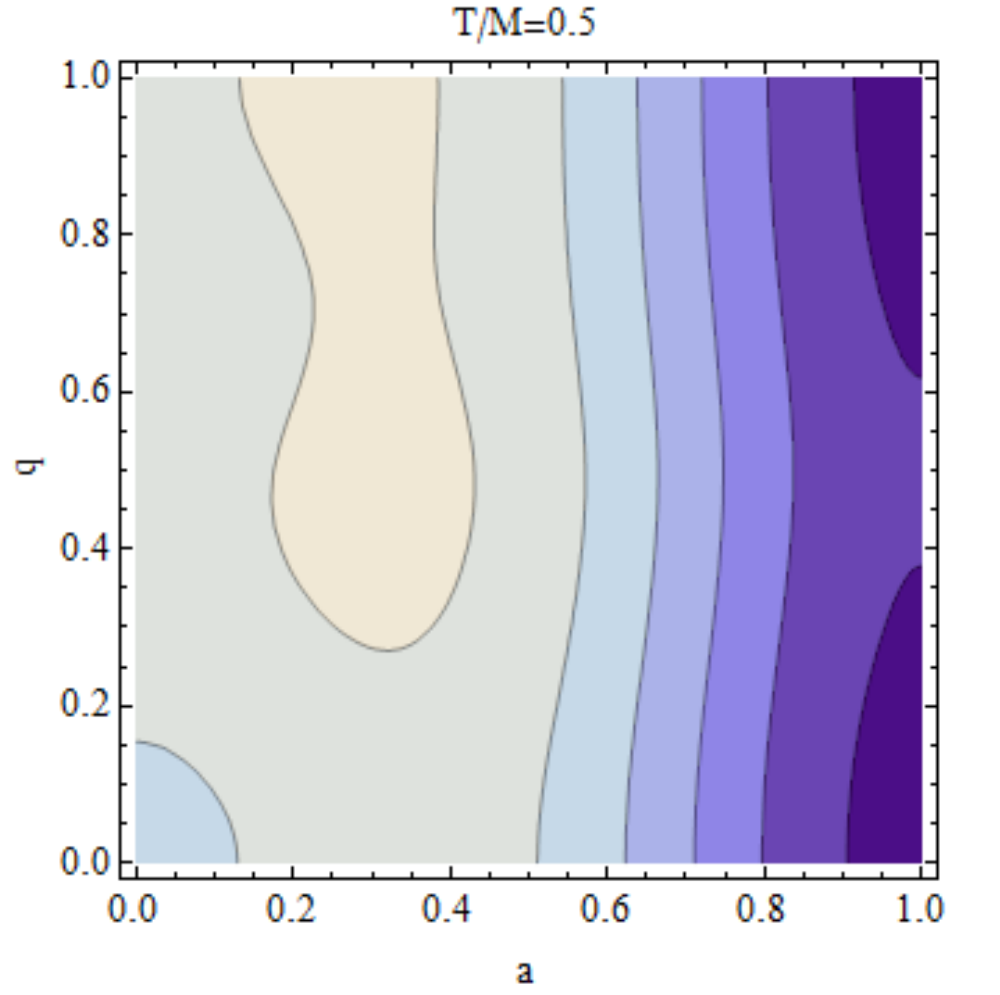}\hspace{.5cm}
\includegraphics[scale=.5]{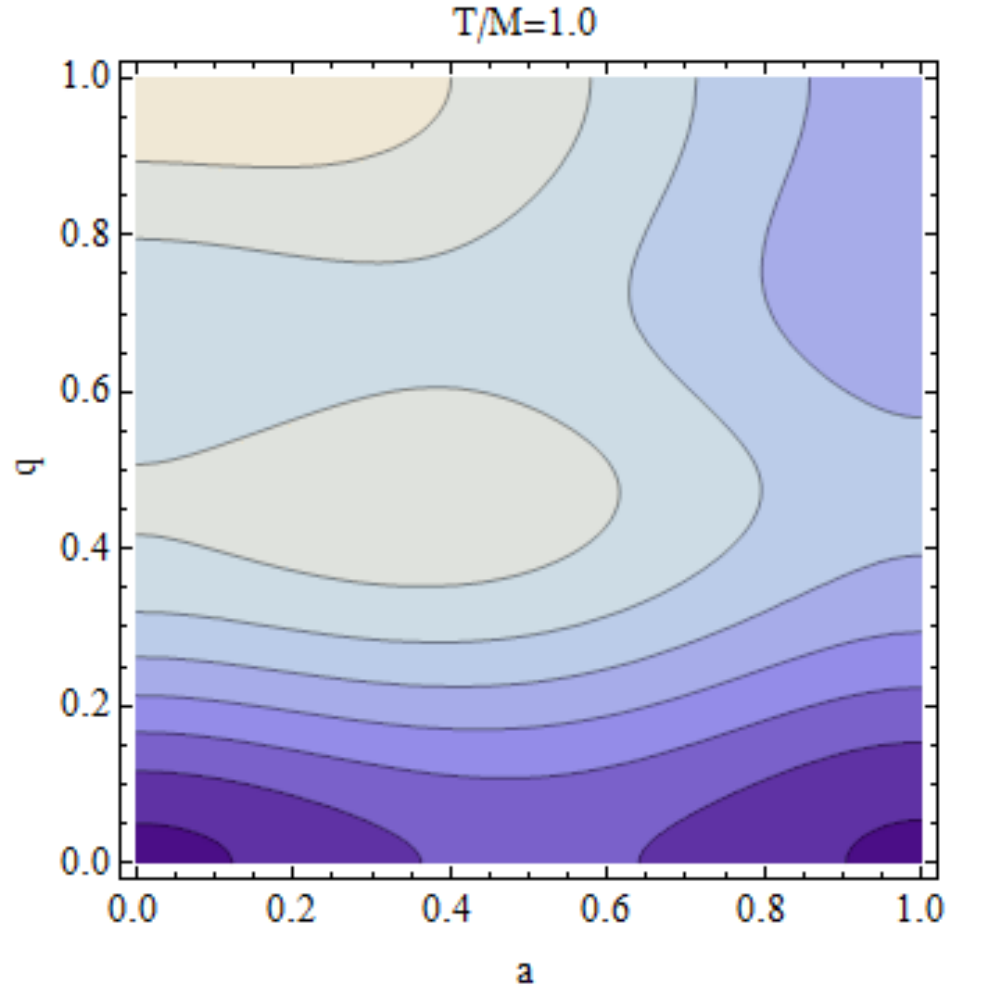}\hspace{.5cm}
\includegraphics[scale=.5]{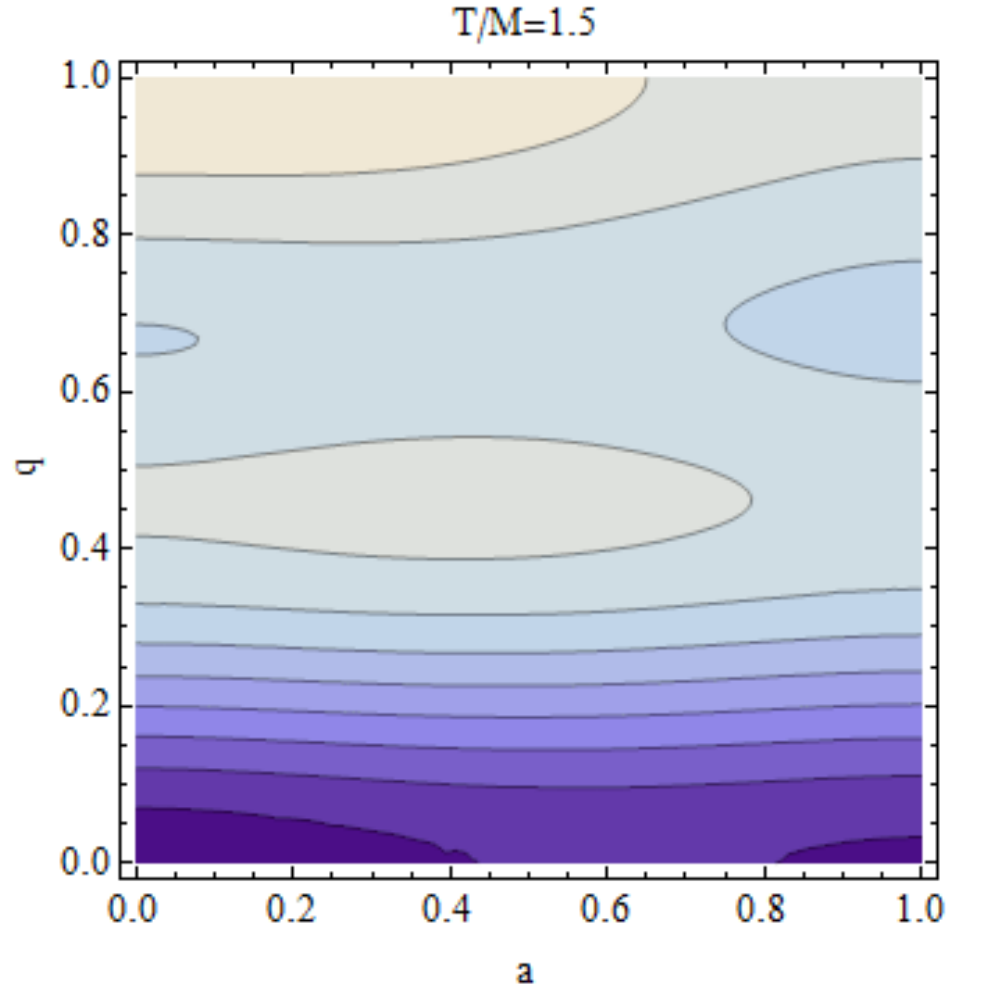}
\caption{The effective potential (\ref{eq:eff_pot_fd+}) with fundamental fermions is shown as contour plots in terms of $a$ and $q$ at several temperatures $T/M=0.5$, $1.0$, and $1.5$.  }
\label{fig:eff_pot_fd+T=00}
\end{figure}

Let us add other matter contents in order to realize the symmetry breaking at low temperatures and observe its restoration at high temperatures. 
In order to reveal the role of each representational field, we study effective potentials of gauge theories by adding fundamental, sextet, and adjoint Dirac fermions separately. 

\begin{figure}[t]
\centering
\includegraphics[scale=.6]{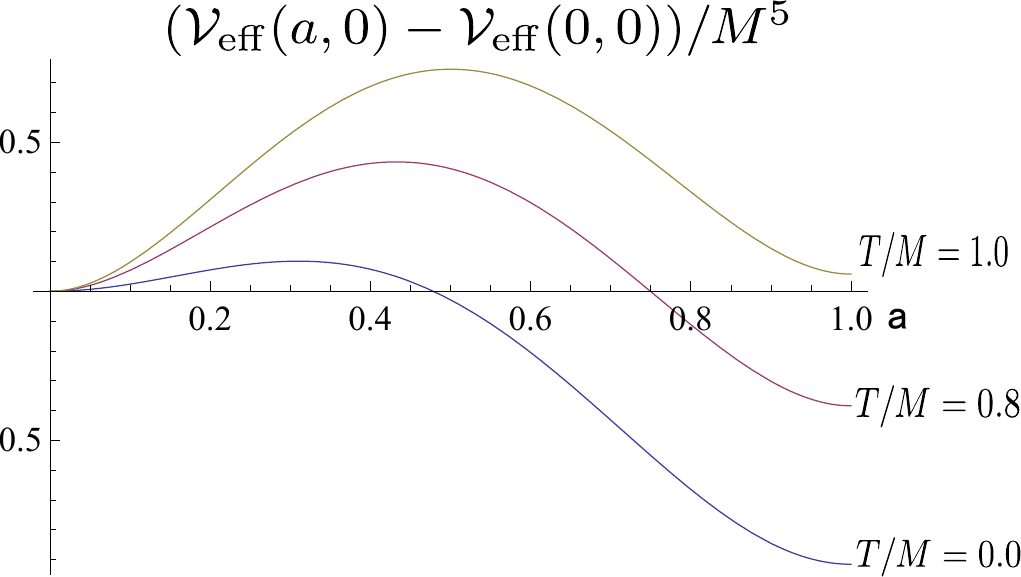}\hspace{.1cm}
\includegraphics[scale=.5]{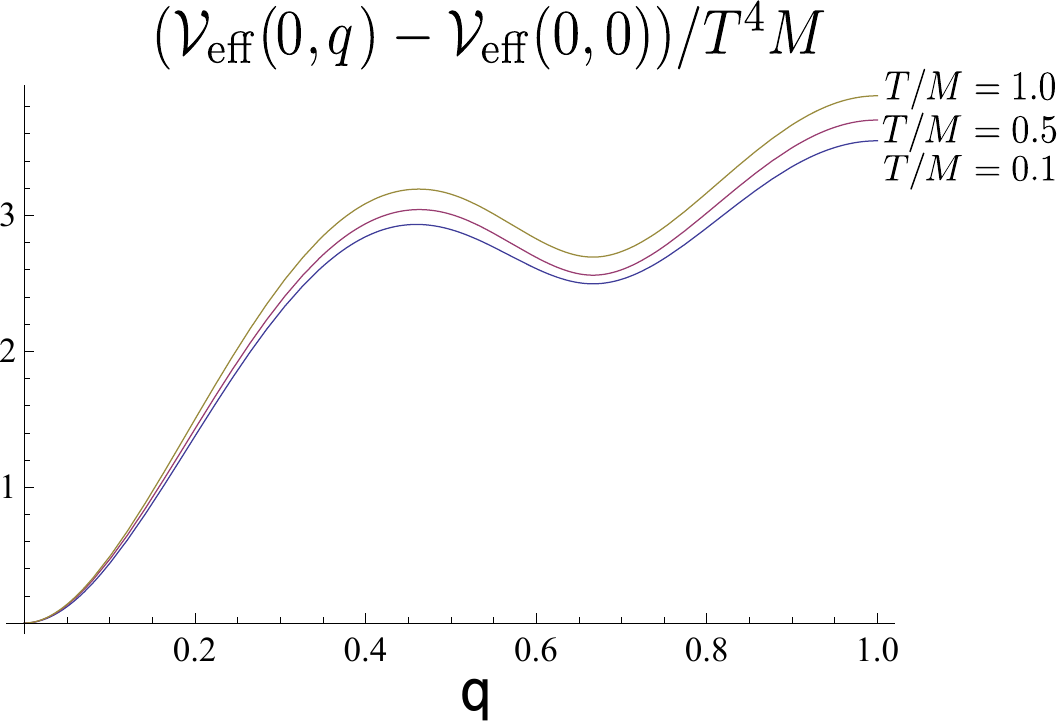}\hspace{.1cm}
\includegraphics[scale=.5]{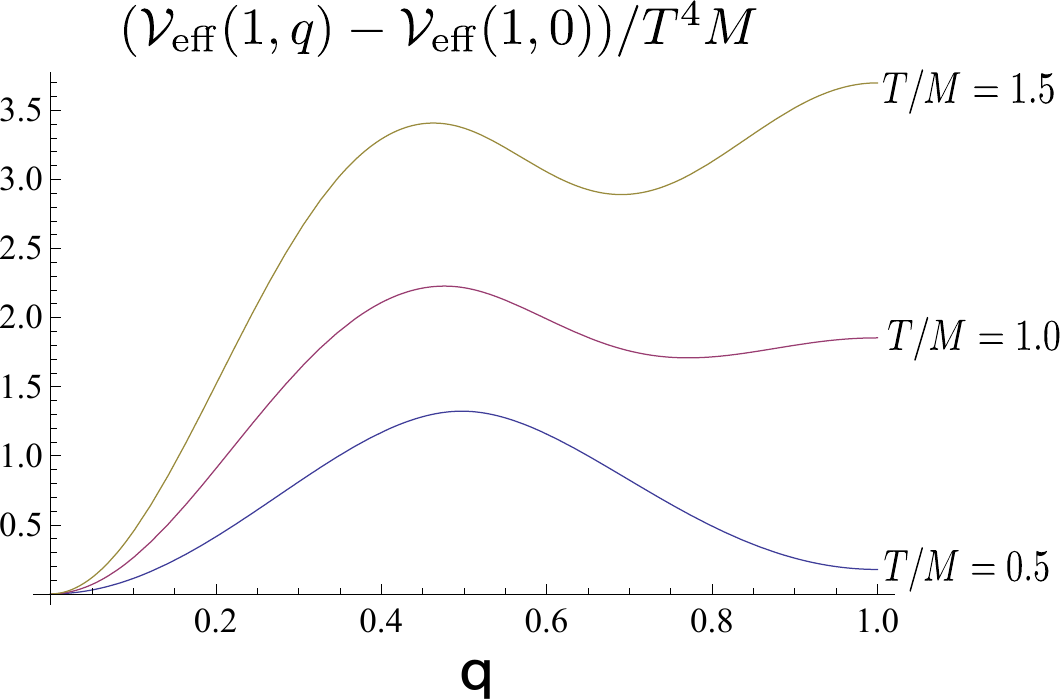}
\caption{Details about the effective potential (\ref{eq:eff_pot_fd+}) for the gauge theory with fundamental fermions. 
 The left panel shows it as a function of $a$ at $q=0$. The center and right panels do as a function of $q$ at $a=0$ and $1$, respectively. }\label{fig:eff_pot_fd_2dim}
\end{figure}

First, we study the effect of fundamental fermions on the effective potential. If the number of fermions is set as $N_F^{\mathrm{fd}(+)}=3$, then the effective potential becomes 
\be
{\cal V}_{\mathrm{eff}}(a,q)=3F_T^{\mathrm{adj}}(q,a,0,0)-12F_T^{\mathrm{fd}}(q,a,1,0). 
\label{eq:eff_pot_fd+}
\ee
At $T=0$, this theory shows $U(1)\times U(1)$ gauge symmetry due to the
nontrivial vacuum (see Fig.~\ref{fig:eff_pot_fd+T=00}). Therefore, fundamental fermions with $\eta\eta'=+1$ stabilizes the point $(a,q)=(1,0)$ at sufficiently low temperatures. 
Since fundamental fermions explicitly break the center symmetry, the symmetry under $q\mapsto q+{2\over 3}$ no longer exists. 
However, some metastable minima are observed along the $q$ direction:
according to the center panel of Fig.~\ref{fig:eff_pot_fd_2dim}, metastable minima exist around $(a,q)=(0,0)$ and $(0,\pm2/3)$ at any temperatures, which are possibly related by a remnant of the broken center symmetry. 
We can also observe in the right panel of Fig.~\ref{fig:eff_pot_fd_2dim} that there exists a metastable minimum at $(a,q)=(1,1)$ at low temperatures, whose energy is almost degenerate to that of the real vacuum $(a,q)=(1,0)$. 
As a temperature is increased, this approximate degeneracy is solved and the position of the metastable minimum comes closer to $(a,q)=(1,\pm2/3)$. 

At high temperatures, the stable minimum becomes $(a,q)=(0,0)$ and the gauge symmetry is restored to $SU(2)\times U(1)$. 
This restoration of the gauge symmetry is a first order phase transition, which can be observed from contour plots in Fig. \ref{fig:eff_pot_fd+T=00} and also in the left panel of Fig. \ref{fig:eff_pot_fd_2dim}. 

\begin{figure}[t]
\centering
\includegraphics[scale=.7]{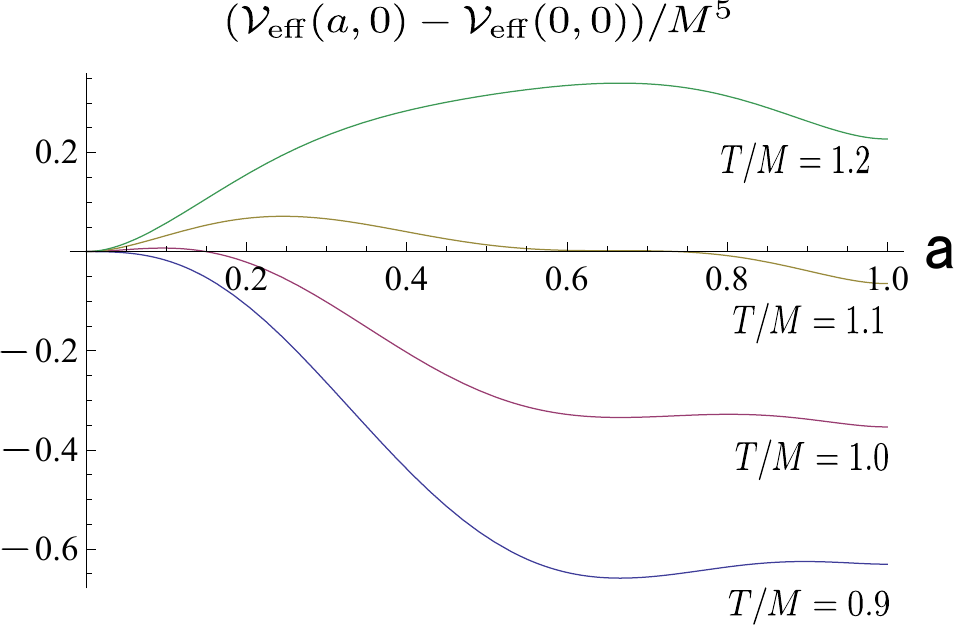}\hspace{.2cm}
\includegraphics[scale=.5]{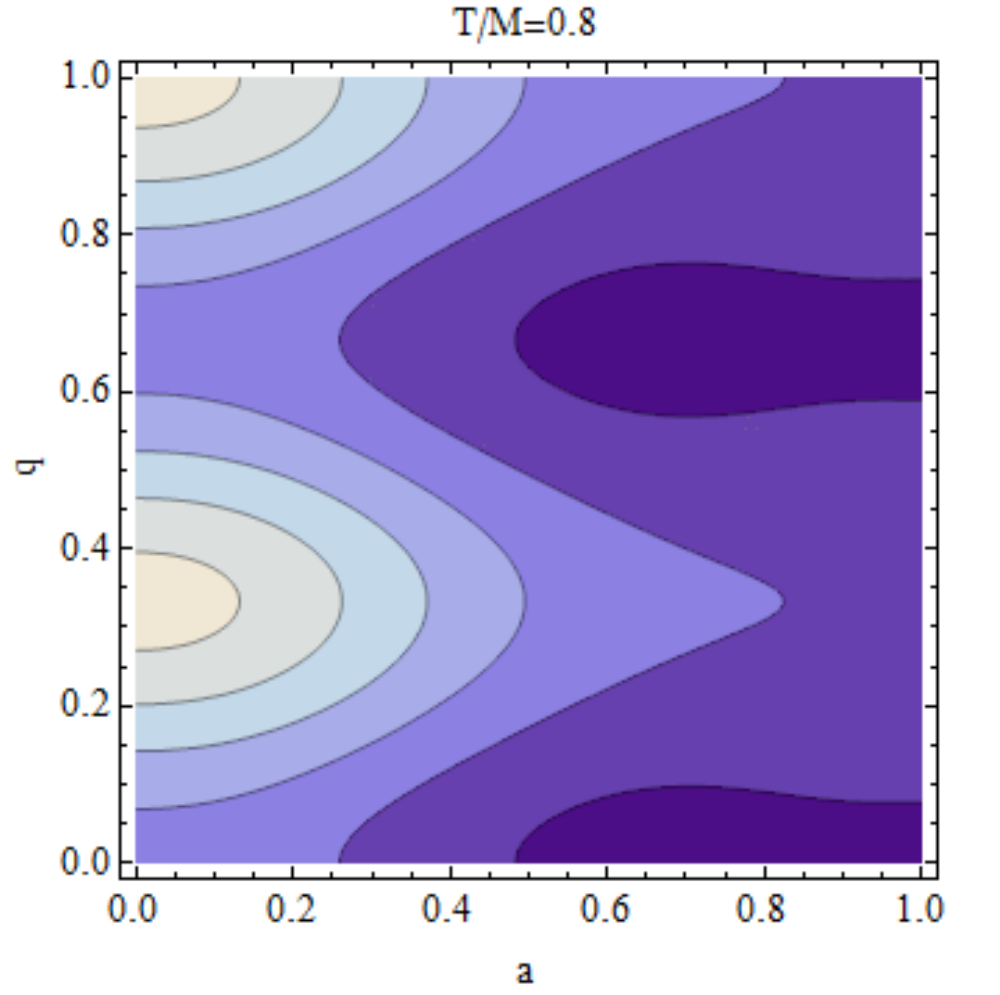}\hspace{.2cm}
\includegraphics[scale=.5]{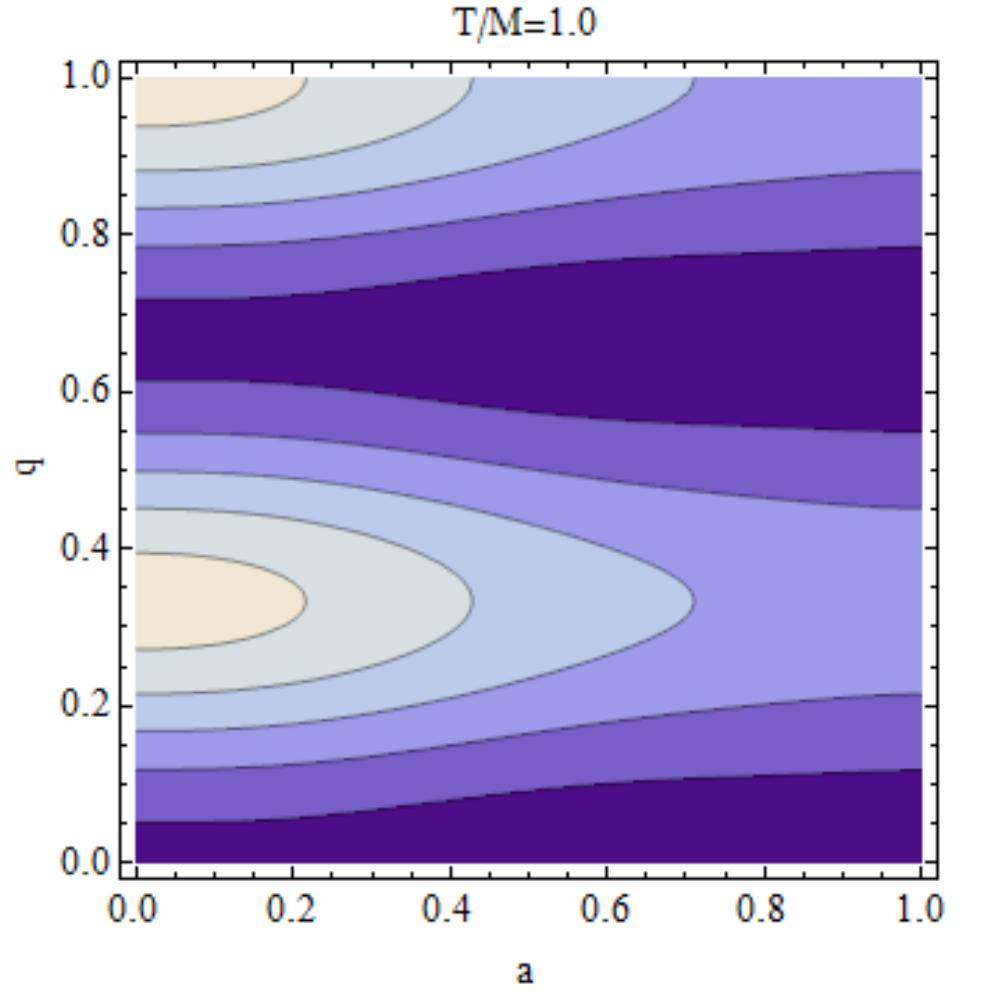}\hspace{.2cm}
\caption{Effective potential (\ref{eq:eff_pot_adj+}) with adjoint fermions along $q=0$, and contour plots at $T/M=0.8$ and $1$.}
\label{fig:eff_pot_adj+}
\end{figure}

To study the effect of adjoint fermions, we put $N_F^{\mathrm{adj}(+)}=2$ and set zero for others. The effective potential of this gauge theory becomes 
\be
{\cal V}_{\mathrm{eff}}(a,q)=3F_T^{\mathrm{adj}}(q,a,0,0)-8F_T^{\mathrm{adj}}(q,a,1,0). 
\label{eq:eff_pot_adj+}
\ee
In Fig.~\ref{fig:eff_pot_adj+}, plots of this effective potential (\ref{eq:eff_pot_adj+}) are shown along $q=0$ for $T/M=0.9$, $1.0$, $1.1$, and $1.2$, and contour plots are also shown for the case $T/M=0.8$ and $1.0$. 
If  temperature is sufficiently high, the potential minimum exists at $(a,q)=(0,0)$, $(0,2/3)$, and $(0,4/3)$, and the system is symmetric. 
This degeneracy is again a consequence of the center symmetry $\mathbb{Z}_3$, since adjoint fermions preserve it. 
As the temperature is lowered so that $T/M\sim 1$, there is a first order phase transition to the $U(1)\times U(1)$ gauge theory. If the temperature is further lowered, there is another first order phase transition to the $U(1)$ gauge theory. This phenomenon is observed also in the previous study of this system without $A_0$ condensation in Ref.~\cite{Maru:2005jy}.


\begin{figure}[t]
\centering
\includegraphics[scale=.55]{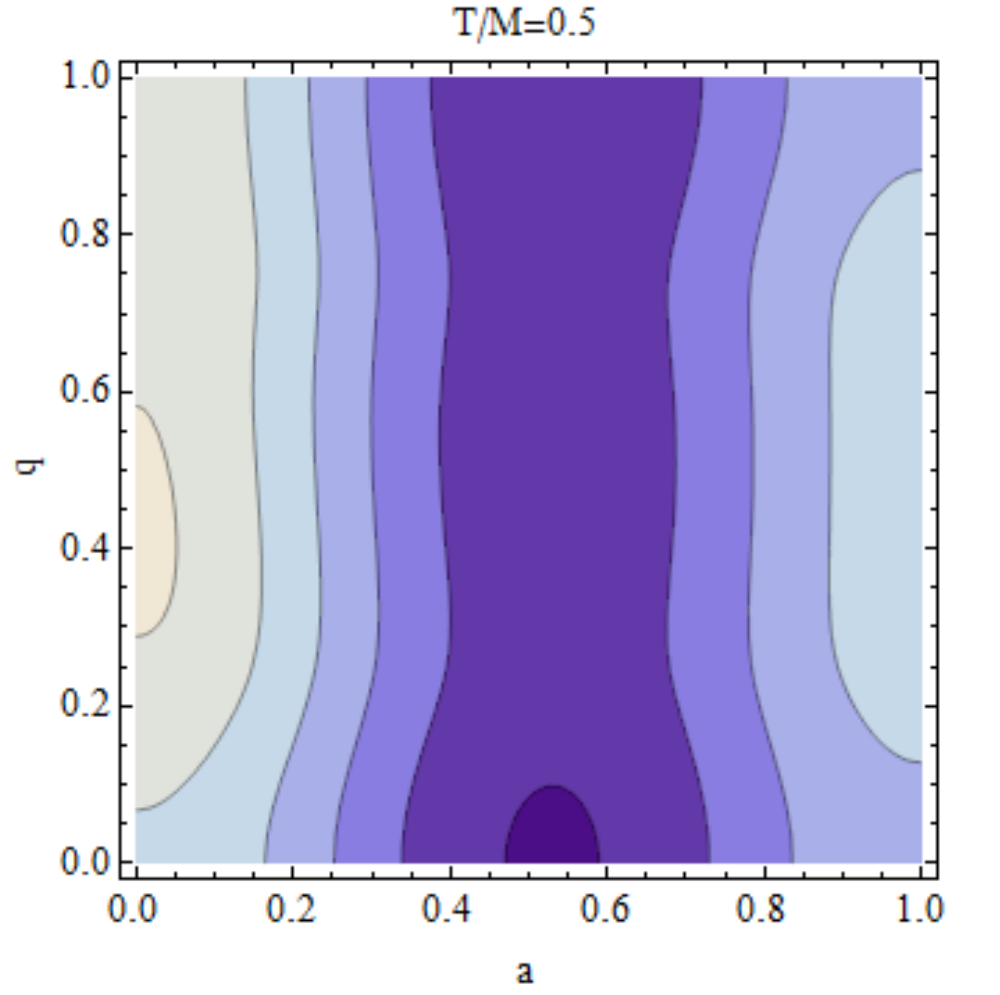}\hspace{.1cm}
\includegraphics[scale=.55]{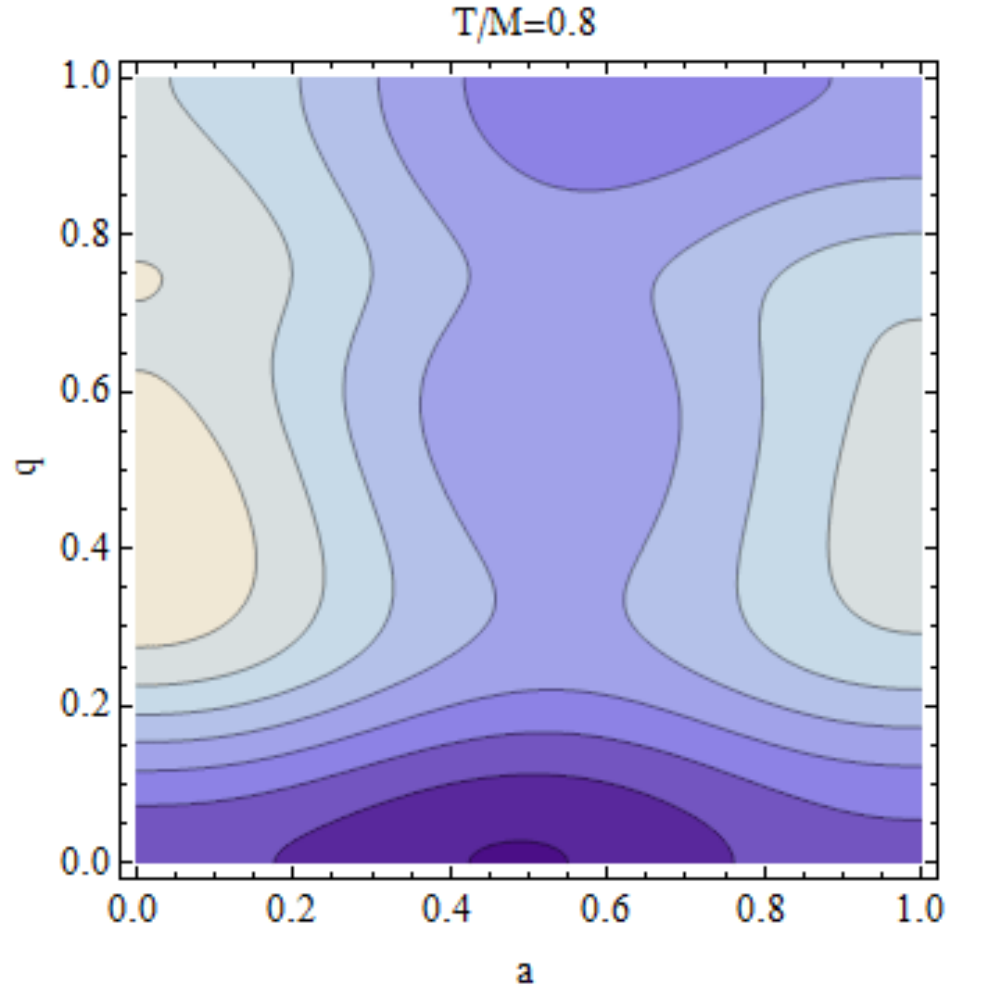}\hspace{.1cm}
\includegraphics[scale=.55]{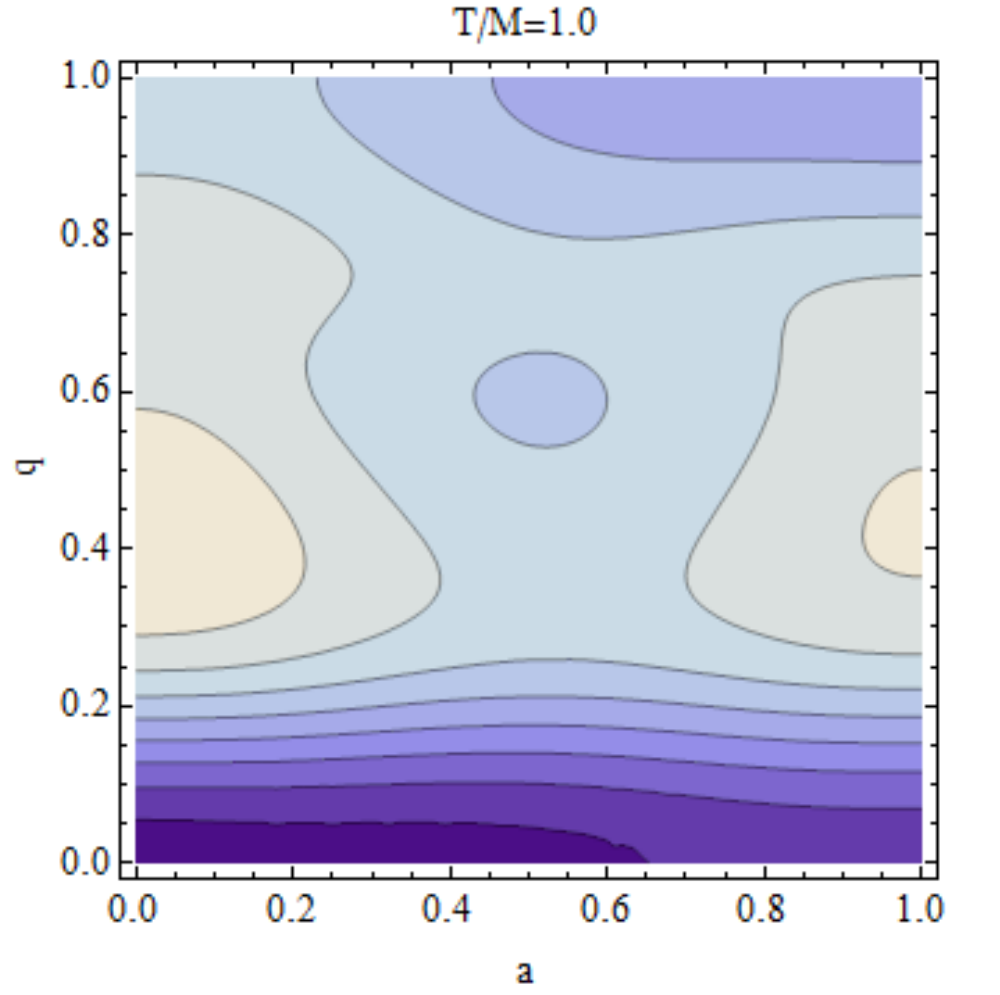}\hspace{.1cm}
\caption{Contour plots of the effective potential (\ref{eq:eff_pot_six+}) with sextet fermions at $T/M=0.5$, $0.8$, and $1.0$. }
\label{fig:eff_pot_six+}
\end{figure}
\begin{figure}[t]
\centering
\includegraphics[scale=.7]{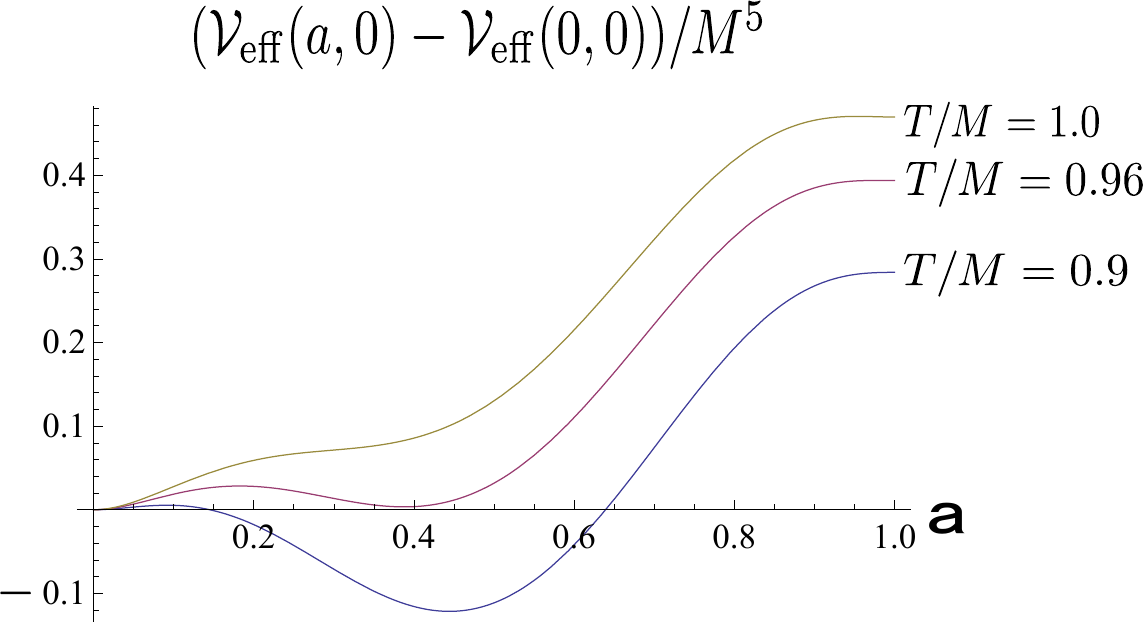}\hspace{.4cm}
\includegraphics[scale=.7]{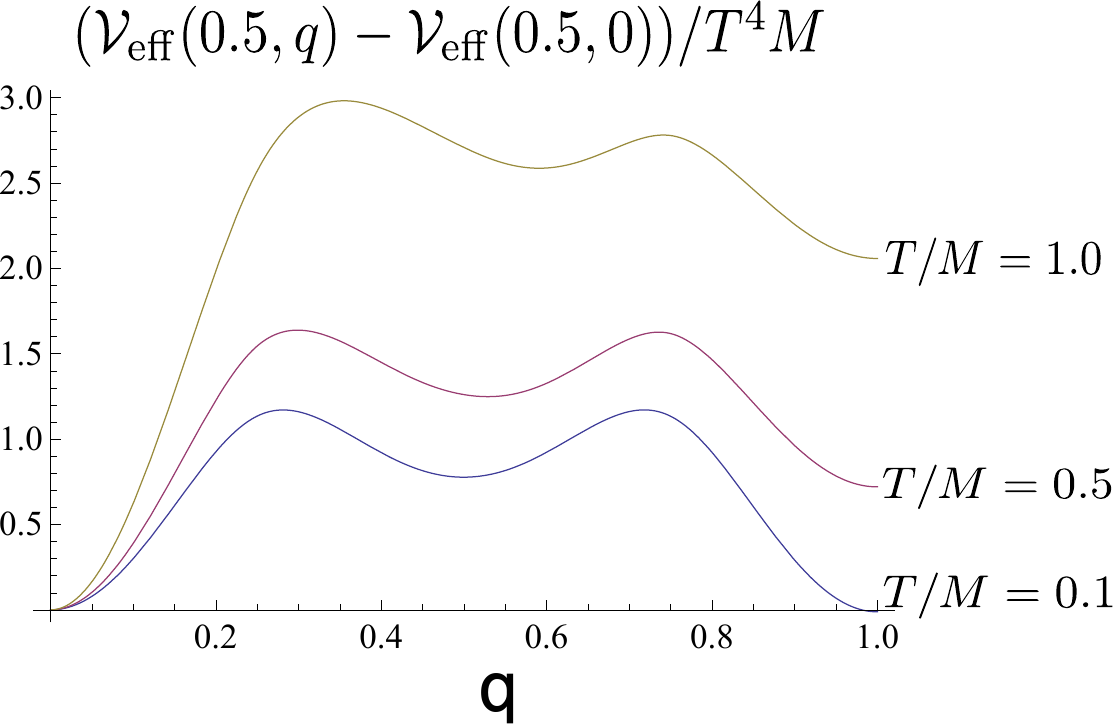}
\caption{Details about the effective potential (\ref{eq:eff_pot_six+}) for the gauge theory with sextet representational fermions. }\label{fig:eff_pot_six+_onedim}
\end{figure}

Finally, we also consider a five-dimensional gauge theory with sextet fermions. The effective potential is given by 
\be
{\cal V}_{\mathrm{eff}}(a,q)=3F_T^{\mathrm{adj}}(q,a,0,0)-8F_T^{\mathrm{sxt}}(q,a,1,0). 
\label{eq:eff_pot_six+}
\ee
Around $T/M=0.96$, the gauge symmetry $SU(2)\times U(1)$ is spontaneously broken to
$U(1)$, and its transition is of the first order (see
Fig.~\ref{fig:eff_pot_six+} and also the left panel of Fig.~\ref{fig:eff_pot_six+_onedim}). 
Again, we can observe the appearance of metastable region along $q$
direction: At low temperatures, the metastable minima lie at $q=0.5$ and $q=1$ , according to the right panel of Fig.~\ref{fig:eff_pot_six+_onedim}. 
This metastable states still survive at $T/M\simeq 1$, although the
original stable state disappears soon after the phase transition as we can
see in the left panel of Fig.~\ref{fig:eff_pot_six+_onedim}. 
Furthermore, at really low temperatures $T/M\lesssim 0.1$, we can observe that the stable and metastable states at $q=0$ and $1$ become almost degenerate.

\section{Summary}\label{sec:summary}
In this paper, we investigated the phase structure of five-dimensional $SU(3)$ gauge theories at finite temperatures. 
In order to study the effect of the temporal gauge-field condensation $\langle A_0\rangle$ and the extra-dimensional gauge-field condensation $\langle A_y\rangle$, we extended the computational technique for the one-loop effective potential at finite temperature on the orbifold $S^1/\mathbb{Z}_2$. 
Importance of the  $A_y$ condensation has been recognized so far in order to describe the electroweak phase transition; however the $A_0$ condensation also turns out to affect the phase structure. 
Indeed, we found new metastable and stable states with nonzero $A_0$ condensations by computing the one-loop effective potential. 
The effect of the UV cutoff is studied in this computation, so that one can check reliability and predictability of these nonrenormalizable five-dimensional gauge theories. 
We introduce the three-dimensional momentum cutoff $\Lambda$ in order to respect the residual gauge symmetry of the effective potential, and find that its effect on Wilson loops vanishes as long as $\Lambda$ is at least about ten times larger than the Kaluza-Klein mass and temperatures. 


Effective potentials with two condensations $\langle A_0\rangle$ and $\langle A_y\rangle$ are shown for simple matter contents, including fundamental, sextet, and adjoint representational Dirac fields. 
So far, fundamental and adjoint fermions has been extensively studied, but sextet fermions can also lead a natural phase transition pattern at finite temperatures. 
For the pure five-dimensional Yang-Mills theory and also for theories only with adjoint fermions, there are several degenerate vacua connected by the center symmetry $\mathbb{Z}_3$ of the original gauge group $SU(3)$. 
In the case of fundamental and sextet fermions, this center symmetry is explicitly broken. However, there still exist metastable minima along the $A_0$ direction, some of which are almost degenerate to the stable state at low temperatures.

These newly found metastable states may play a significant role in dynamics of phase transitions. 
Recently, a domain structure of the deconfined QCD matter has been discussed as a possible scenario to explain properties of the quark-gluon plasma such as the large opacity and the ideal fluidity \cite{Asakawa:2012yv}.
Also in Ref.~\cite{Kashiwa:2013qma},  properties of  quark-gluon plasma are discussed when there are domain structures induced by metastable states.
From our calculation, domain structures in gauge-Higgs unification models are expected to exist thanks to many metastable minima of effective potentials.
If these metastable states also appear in nonequilibrium systems, they could affect the cosmological phase transition \cite{KorthalsAltes:1994be,PhysRevLett.75.2799,Machtey:2009wu}. 

Before closing the summary, let us mention details about the symmetry breaking pattern of each matter content. 
At sufficiently low temperatures, Dirac fermions contribute to the spontaneous breaking of the gauge symmetry $SU(2)\times U(1)$. 
For sextet and adjoint fermions, the gauge symmetry is broken to $U(1)$ but the way of its restoration is very different. 
In the case of theories with sextet fermions, there is a restoration to the $SU(2)\times U(1)$ gauge theory as a first order phase transition. However, for theories with adjoint fermions, they once are restored to $U(1)\times U(1)$, and after that the $SU(2)\times U(1)$ gauge theory appears. 
For theories with fundamental fermions, the symmetry is spontaneous broken to $U(1)\times U(1)$ at sufficiently low temperatures, and it is restored to $SU(2)\times U(1)$ under first-order phase transition. 

In this paper, we consider the case of zero fermion number density. It must be an interesting task to investigate the phase structure at finite fermion number density, and, for that purpose, we need to introduce chemical potentials for conserved charges. 
Since we included the effect of $\langle A_0\rangle$, we would like to emphasize that this formalism can be extended to the case with finite chemical potentials in a straightforward way for studying finite density systems. 
In this case, however, the perturbative effective potential can acquire
its imaginary part as is known in the case of quantum chromodynamics (see Ref.~\cite{Fukushima:2006uv} for example). Thus, we must overcome this difficulty of the sign problem even at perturbative calculations. 

\appendix
\section{Background field gauge}\label{app:back}
We start from the original five-dimensional Yang-Mills action, 
\be
S_{\mathrm{YM}}=\int \mathrm{Tr} [F_A\wedge * F_A], 
\ee
with $F_A=\diff A+i g A\wedge A$ and $A=A_I^a T^a \diff x^I$. We decompose the gauge field into two parts, $A=A^{\mathrm{cl}}+A^{\mathrm{qu}}$, where $A^{\mathrm{cl}}$ is a background field and $A^{\mathrm{qu}}$ describes the quantum fluctuation. 
For our purpose, the classical field strength $F_{A^{\mathrm{cl}}}$ can be assumed to be zero, and thus the covariant derivative $D^{\mathrm{cl}}_{I}=\p_I +ig A^{\mathrm{cl}}_I$ commutes to each other (for any representations). The field strength becomes $F_{A}=D^{\mathrm{cl}} A^{\mathrm{qu}}+ig A^{\mathrm{qu}}\wedge A^{\mathrm{qu}}$. 

The second order term of the Yang-Mills action in terms of $A^{\mathrm{qu}}$ is given by 
\be
S^{(2)}_{\mathrm{YM}}=\int \mathrm{Tr}\left[(D^{\mathrm{cl}}A^{\mathrm{qu}})\wedge *(D^{\mathrm{cl}}A^{\mathrm{qu}})\right]=\int \diff^4 x\diff y\; \mathrm{Tr}\left[(D^{\mathrm{cl}}_I A^{\mathrm{qu}}_J)(D^{\mathrm{cl}}_I A^{\mathrm{qu}}_J-D^{\mathrm{cl}}_J A^{\mathrm{qu}}_I)\right], 
\ee
with $D^{\mathrm{cl}}_I A^{\mathrm{qu}}_J=\p_I A^{\mathrm{qu}}_J+ig[A^{\mathrm{cl}}_I,A^{\mathrm{qu}}_J]$. We take the gauge fixing function as 
\be
D_I^{\mathrm{cl}}A^{\mathrm{qu}}_I=0, 
\ee
and the classical background field $A^{\mathrm{cl}}$ must also satisfy this condition. Adding the Faddeev-Popov ghost fields $c$ and $\overline{c}$, we find that the quadratic term of the action becomes 
\be
S^{(2)}_{\mathrm{YM+FP}}=\int\diff^4 x\diff y \; \mathrm{Tr}\left[ (D^{\mathrm{cl}}_IA^{\mathrm{qu}}_J)^2+\overline{c}D^{\mathrm{cl}}_ID^{\mathrm{cl}}_I c \right]. 
\ee

\section{Calculations for the effective potential at finite temperatures}\label{app:cal}
In order to calculate the effective potential, we need to evaluate the following quantity, 
\be
F_T( q ,a)={1\over 2}\int_{0}^{\Lambda}{\diff^3\bm{p}\over (2\pi)^3}TM\sum_{n,m=-\infty}^{\infty}
\ln\left[{\bm{p}^2+(2\pi T)^2(n+ {q\over 2} )^2+(2\pi M)^2 (m+{a\over 2})^2 \over \bm{p}^2+(2\pi T)^2 n^2+(2\pi M)^2 m^2}\right],  
\ee
with $T$ the temperature, and $M$ the inverse radius of the compact dimension. 
$\Lambda$ is the UV cutoff for the spatial momentum. In this expression, $n$ and $m$ represent the Matsubara and Kaluza-Klein modes, respectively. Using the proper time method, we can perform the spatial momentum integration: 
\bea
F_T( q ,a)&=&{TM\over 2}\sum_{n,m\in\mathbb{Z}}\int_0^{\Lambda}{\diff^3\bm{p}\over (2\pi)^3}
\left[-\int_0^{\infty}{\diff s\over s}e^{-s\bm{p}^2}\right.\nonumber\\
&&\times\left.\left(e^{-s((2\pi T)^2(n+ q/2 )^2+(2\pi M)^2(m+a/2)^2)}-e^{-s((2\pi T)^2n^2+(2\pi M)^2 m^2)}\right)\right]\\
&=&{TM\over 2(4\pi)^{3/2}}\int_{0}^{\infty}{\diff s\over s^{5/2}}\left(\mathrm{Erf}(\Lambda\sqrt{s})-{2\over \sqrt{\pi}}\Lambda\sqrt{s}e^{-\Lambda^2 s}\right)\nonumber\\
&&\times \left[-\sum_{n,m\in\mathbb{Z}}\left(e^{-s((2\pi T)^2(n+ q/2 )^2+(2\pi M)^2(m+a/2)^2)}-e^{-s((2\pi T)^2n^2+(2\pi M)^2 m^2)}\right)\right].
\eea
Using Poisson's resummation formula, we find that 
\bea
F_T(q,a)&=& {TM\over 2(4\pi)^{3/2}}\int_{0}^{\infty}{\diff s\over s^{5/2}}\left(\mathrm{Erf}(\Lambda\sqrt{s})-{2\over \sqrt{\pi}}\Lambda\sqrt{s}e^{-\Lambda^2 s}\right)\nonumber\\
&&\times \sum_{\tn,\tm\in\mathbb{Z}}\sqrt{1\over 4\pi M^2 s}\sqrt{1\over 4\pi T^2 s}
e^{-{\tm^2\over 4M^2 s}}e^{-{\tn^2\over 4T^2 s}}\left(1-e^{-\pi i(\tn  q +\tm a)}\right)\\
&=&{ M^5\over 2\pi^{5/2}}\sum_{\tn,\tm\in\mathbb{Z}}{1-e^{-\pi i(\tn  q +\tm a)}\over (\tm^2+{M^2\over T^2}\tn^2)^{5/2}}\times\int_{0}^{\infty}\diff \tau \tau^{5/2-1}e^{-\tau}
\nonumber\\
&&\times\left(\mathrm{Erf}\left({\Lambda\over 2M}{\sqrt{\tm^2+{M^2\over T^2}\tn^2}\over \sqrt{\tau}}\right)-{\Lambda\over \sqrt{\pi}M}{\sqrt{\tm^2+{M^2\over T^2}\tn^2}\over \sqrt{\tau}}e^{-{\Lambda^2\over 4M^2}{\tm^2+{M^2\over T^2}\tn^2\over \tau}}\right). 
\label{app_eq01}
\eea
When $\tm =0$ and $\tn=0$, the summands are independent of $a$ and $q$, which are not of our interest. In the following, the summation over $\tm$ and $\tn$ is restricted to $\tm\not=0$ or $\tn\not=0$. 

The integration in (\ref{app_eq01}) can be done explicitly to get 
\bea
F_T(q,a)&=&{M^5\over 2\pi^3}\sum_{\tn,\tm\in\mathbb{Z}}{1-e^{-\pi i(\tn  q +\tm a)}\over (\tm^2+{M^2\over T^2}\tn^2)^{5/2}}\left[G^{2,1}_{1,3}\left(\begin{array}{ccc}&1&\\{1\over 2}&{5\over 2}&0\end{array};{\Lambda^2\over 4M^2}(\tm^2+{M^2\over T^2}\tn^2)\right)\right.\nonumber\\
&&\left.-4\left({\Lambda\over 2M}\sqrt{\tm^2+{M^2\over T^2}\tn^2}\right)^3K_2\left({\Lambda\over M}\sqrt{\tm^2+{M^2\over T^2}\tn^2}\right)\right].
\eea
Here, $G$ is the Meijer $G$ function defined by 
\be
G^{m,n}_{p,q}\left(\begin{array}{ccc}a_1&\ldots&a_p\\b_1&\ldots&b_q\end{array};z\right)=\int_L{\diff s\over 2\pi i}z^s {\prod_{j=1}^{m}\Gamma(b_j-s) \prod_{j=1}^{n}\Gamma(1-a_j+s)\over \prod_{j=m+1}^{q} \Gamma(1-b_j+s)\prod_{j=n+1}^{p}\Gamma(a_j-s)}, 
\ee
where $L$ is an upward oriented loop contour which separates the poles of $\prod_{j=1}^{m}\Gamma(b_j-s)$ from those of $\prod_{j=1}^{n}\Gamma(1-a_j+s)$ and which begins and ends at $+\infty$ \cite{Askey2010}. 
Since this quantity is convergent in the limit $\Lambda\to \infty$, we separate it into two parts by its $\Lambda$ dependence: 
\be
F_{T}( q ,a)=F_T( q ,a)|_{\Lambda\mathrm{-indep.}}+F_T( q ,a)|_{\Lambda\mathrm{-dep.}}.  
\ee
The $\Lambda$-independent part is defined by taking the limit $\Lambda\to\infty$ of the field-dependent part of the effective potential: 
\bea
F_T( q ,a)|_{\Lambda\mathrm{-indep.}}&=&{3M^5\over 4\pi^2}\sum_{\tm\ge 1}{1-\cos \pi \tm a\over \tm^5}+{3M^5\over 4\pi^2}\sum_{\tn\ge 1}{1-\cos \pi \tn  q \over \sqrt{{M^2\over T^2}\tn^2}^5}\nonumber\\
&&+{3M^5\over 2\pi^2}\sum_{\tm,\tn\ge 1}{1-\cos \pi \tn  q  \cos \pi \tm a\over \left(\tm^2+{M^2\over T^2}\tn^2\right)^{5/2}}. 
\eea
On the other hand, by subtracting the above UV-finite field-dependent part, we can obtain the cutoff dependence of the effective potential as follows: 
\bea
F_T(q,a)|_{\Lambda\mathrm{-dep.}}&=&{M^5\over \pi^3}\sum_{\tm\ge 1}{1-\cos \pi \tm a\over \tm^5}\widetilde{G}\left({\Lambda\over 2M}\tm\right)
+{M^5\over \pi^3}\sum_{\tn\ge 1}{1-\cos \pi \tn  q \over \sqrt{{M^2\over T^2}\tn^2}^5} \widetilde{G}\left({\Lambda\over 2T}\tn\right)\nonumber\\
&&+{2M^5\over \pi^3}\sum_{\tm,\tn\ge 1}{1-\cos \pi \tn  q  \cos \pi \tm a\over \left(\tm^2+{M^2\over T^2}\tn^2\right)^{5/2}}\widetilde{G}\left({\Lambda\over 2M}\sqrt{\tm^2+{M^2\over T^2}\tn^2}\right). 
\eea
\begin{figure}[t]
\centering\includegraphics[scale=0.5]{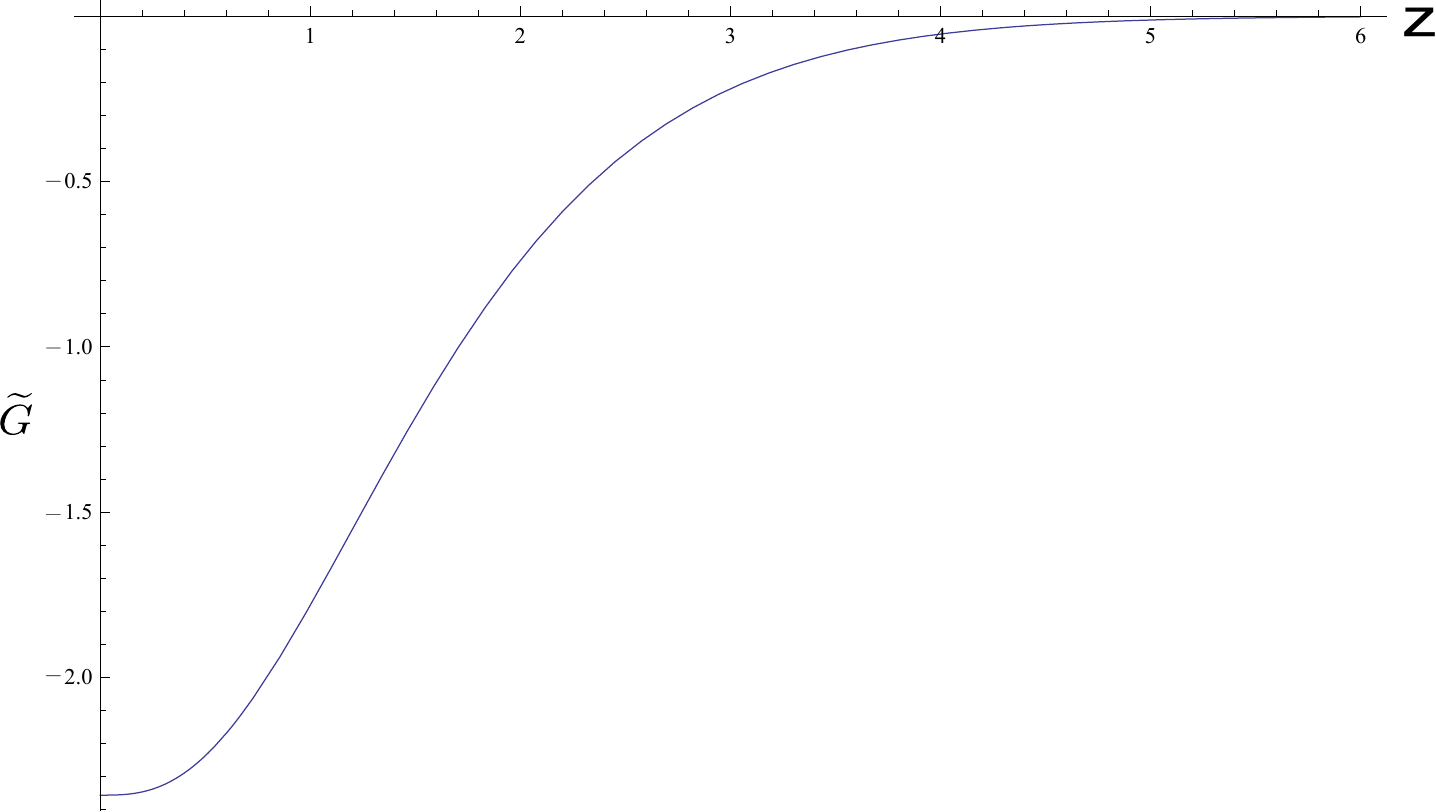}
\caption{Behavior of the special function $\widetilde{G}$. }
\label{fig:special}
\end{figure}
Here we define a special function 
\be
\widetilde{G}(z)=G^{2,1}_{1,3}\left(\begin{array}{ccc}&1&\\{1\over 2}&{5\over 2}&0\end{array};z^2\right)-{3\pi \over 4}-4z^3 K_2(2z). 
\ee
Behavior of $\widetilde{G}$ along the positive real axis is shown in
Fig.~\ref{fig:special}, and it vanishes exponentially fast when $z\gtrsim 5$. Therefore, the result becomes independent of the three-dimensional UV cutoff as long as $\Lambda\gtrsim 10\mathrm{max}\{M, T\}$.

\section{Lie algebra $\mathfrak{su}(3)$}\label{app:lie}
In order to make a firm connection between charges of fields in the gauge-Higgs unification model and the $\mathfrak{su}(3)$ Lie algebra, we take the basis by half of the Gell-Mann matrices: 
\bea
&&T^1={1\over 2}\left(\begin{array}{ccc}0&1&0\\1&0&0\\0&0&0\end{array}\right),\quad 
T^2={1\over 2}\left(\begin{array}{ccc}0&-i&0\\ i&0&0\\0&0&0\end{array}\right),\quad
T^3={1\over 2}\left(\begin{array}{ccc}1&0&0\\0&-1&0\\0&0&0\end{array}\right),\nonumber\\
&&T^4={1\over 2}\left(\begin{array}{ccc}0&0&1\\0&0&0\\1&0&0\end{array}\right),\quad
T^5={1\over 2}\left(\begin{array}{ccc}0&0&-i\\0&0&0\\i&0&0\end{array}\right),\nonumber\\
&&T^6={1\over 2}\left(\begin{array}{ccc}0&0&0\\0&0&1\\0&1&0\end{array}\right),\quad 
T^7={1\over 2}\left(\begin{array}{ccc}0&0&0\\0&0&-i\\0&i&0\end{array}\right),\quad  
T^8={1\over 2\sqrt{3}}\left(\begin{array}{ccc}1&0&0\\0&1&0\\0&0&-2\end{array}\right). 
\eea
Under the inner product of $\mathfrak{su}(3)$, this basis is normalized as $(T^i,T^j)\equiv \mathrm{Tr}(T^i T^j)=\delta^{ij}/2$, 

Since we take the direction of the gauge field condensation $\langle A_6\rangle$ along $T^6$ in (\ref{eq:background}), it is appropriate to take the Cartan subalgebra as $\mathcal{H}=\mathrm{span}\{T^6,{\sqrt{3}\over 2}(T^3+T^8/\sqrt{3})\}$. 
Let us denote this basis as $H_1=T^6$ and $H_2={\sqrt{3}\over 2}(T^3+T^8/\sqrt{3})$. 
The simple roots are given by $\alpha_1=({1\over 2},{\sqrt{3}\over 2})$ and $\alpha_2=({1\over 2},-{\sqrt{3}\over 2})$. Corresponding root vectors are 
\be
E_{\alpha_1}={1\over 2}\{(T^1+i T^2)-(T^4+i T^5)\},\quad E_{\alpha_2}={1\over 2}\{(T^1-i T^2)+(T^4-i T^5)\}, \label{eq:simple_root}
\ee
respectively. The last positive root vector  is obtained as 
\be
E_{\alpha_1+\alpha_2}=\sqrt{2}[E_{\alpha_1},E_{\alpha_2}]={1\over 2\sqrt{2}}\{(T^3-\sqrt{3}T^8)+2i T^7\}. \label{eq:last_positive}
\ee
Root vectors with negative roots are given by $E_{-\alpha}=(E_{\alpha})^{\dagger}$. 
This system of root vectors satisfies the normalization $(E_{\alpha},E_{-\alpha})=1/2$ and $[E_{\alpha},E_{-\alpha}]={1\over 2}\alpha\cdot H$. Unless $\alpha+\beta=0$, $(E_{\alpha},E_{\beta})=0$, and $(H_i,E_{\alpha})=0$ in general.

The inverse of (\ref{eq:simple_root}) and (\ref{eq:last_positive}) is given by 
\bea
\begin{array}{ll}
T^1={1\over 2}(E_{\alpha_1}+E_{-\alpha_1}+E_{\alpha_2}+E_{-\alpha_2}), &
T^2={1\over 2i}(E_{\alpha_1}-E_{-\alpha_1}-E_{\alpha_2}+E_{-\alpha_2}), \\
T^3={\sqrt{3}\over 2}H_2+{1\over 2\sqrt{2}}(E_{\alpha_1+\alpha_2}+E_{-\alpha_1-\alpha_2}), & T^8={1\over 2}H_2-{\sqrt{3}\over 2\sqrt{2}}(E_{\alpha_1+\alpha_2}+E_{-\alpha_1-\alpha_2}),\\
T^4={1\over 2}(-E_{\alpha_1}-E_{-\alpha_1}+E_{\alpha_2}+E_{-\alpha_2}), &
T^5={1\over 2i}(-E_{\alpha_1}+E_{-\alpha_1}-E_{\alpha_2}+E_{\alpha_2}), \\
T^6=H_1,&T^7={1\over \sqrt{2}i}(E_{\alpha_1+\alpha_2}-E_{-\alpha_1-\alpha_2}). 
\end{array}
\eea

\begin{acknowledgments}
The authors thank Kazunori Takenaga for useful discussion. 
K.K. is supported by the RIKEN special postdoctoral researchers program and by JSPS Research Fellowships for Young Scientists. 
Y.T. is supported by JSPS Research Fellowships for Young Scientists. 
This work was partially supported by the JSPS Strategic Young Researcher Overseas Visits Program for Accelerating Brain Circulation and by the Program for Leading Graduate Schools, MEXT, Japan.
\end{acknowledgments}

\bibliography{./ref}

\end{document}